\documentclass[aps,twocolumn,superscriptaddress,showpacs]{revtex4}



\usepackage{natbib}
\usepackage{graphicx,epsfig,rotating,color}

\usepackage{tabularx}

\usepackage{graphicx}
\usepackage{dcolumn}

\usepackage{subfigure}

\begin{document}

\title{Core of communities in bipartite networks}

\author{Christian Bongiorno}
\affiliation{Dipartimento di Fisica e Chimica, Universit\`a degli Studi di Palermo, Viale delle Scienze Ed. 18, I-90128, Palermo, Italy}

\author{Andr\'as London}
\affiliation{Institute of Informatics, University of Szeged, \'Arp\'ad t\'er 2, H-6720, Szeged, Hungary}

\author{Salvatore Miccich\`e}
\affiliation{Dipartimento di Fisica e Chimica, Universit\`a degli Studi di Palermo, Viale delle Scienze Ed. 18, I-90128, Palermo, Italy}

\author{Rosario N. Mantegna}
\affiliation{Dipartimento di Fisica e Chimica, Universit\`a degli Studi di Palermo, Viale delle Scienze Ed. 18, I-90128, Palermo, Italy}
\affiliation{Center for Network Science, Central European University, Nador 9, H-1051, Budapest, Hungary}
\affiliation{Department of Computer Science, University College London, Gower Street, London, WC1E 6BT, UK}


\date{\today}

\begin{abstract}
We use the information present in a bipartite network to detect cores of communities of each set of the  bipartite system.  Cores of communities are found by investigating statistically validated projected networks obtained using information present in the bipartite network. Cores of communities are highly informative and robust with respect to the presence of errors or missing entries in the bipartite network. We assess the statistical robustness of cores by investigating an artificial benchmark network, the co-authorship network, and the actor-movie network. The accuracy and precision of the partition obtained with respect to the reference partition are measured in terms of the adjusted Rand index and of the adjusted Wallace index respectively. The detection of cores is highly precise although the accuracy of the methodology can be limited in some cases.
\end{abstract}



\maketitle


\section{Introduction}

Community detection in networks \cite{Fortunato2016,Fortunato2010} is one of the major research areas in network science \cite{Newman2010,Barabasi2016}. Community detection in networks (also called network clustering) is performed with a variety of methods because there are no universal protocols on basic aspects of the problems \cite{Fortunato2016}. It is therefore important to point out aspects of community detection approaches that are informative with respect to the robustness and reproducibility of the results obtained with most popular community detection algorithms.

One of the most popular community detection method is the one based on modularity optimization. Modularity is a quality function introduced in ref. \cite{Newman2004}. Several community detection algorithms maximizing this quality function have been proposed. One of the most widespread ones is the so-called Louvain algorithm \cite{Blondel2008} that is highly efficient in clustering large networks. 

Community detection performed with modularity optimization is relatively simple, practical and efficient 
but it also presents some limitations. In fact, it is well known that modularity optimization presents a resolution limit \cite{Fortunato2007}. Moreover, the approaches of modularity optimization adopting suitable multiresolution versions of it \cite{Reichardt2006,Arenas2008} are in most cases not able to fully solve the problem \cite{Lancichinetti2011a}. In practical cases modularity optimization can detect partitions characterized by very close modularity values,  and these partitions can disagree in the composition of the largest  modules and in the distribution of module size \cite{Good2010}. Several of these partitions associated with degenerate solutions can be poorly correlated the one with each other \cite{Zhang2014}.
 
It is therefore of interest to assess which part of the partitions is more robust with respect to the limitations of the methodology and with respect to the potential unknowns and errors present in real data. Community detection is performed in several types of networks. In most common case all nodes of the network are of the same type and are connected by binary or weighted links. 
Bipartite networks are networks where nodes can be divided into two sets, say A and B, and links connect nodes of the different sets only. In the investigation of bipartite networks, as for example an actor-movie network or an author scientific paper network, the customary approach is to project the bipartite network to obtain a network of nodes of the same type (for example a network of movies in the case of the actor-movie network). Community detection is usually performed in projected networks although it can also be performed in bipartite networks directly \cite{Barber2007,Larremore2014}. The information present in a bipartite network is richer than the information transferred in the two corresponding projected networks. Therefore the investigation of properties of community detection in projected networks originating from a bipartite network can be informative about the reliability and robustness of the partitions obtained. 

Recent studies have considered the statistical reliability of community detection in networks \cite{Karrer2008,Lancichinetti2010,Lancichinetti2011b}.
In this paper we investigate (i) the degree of informativeness, and (ii) the robustness to incompleteness and accuracy of the links of the bipartite network, of partitions obtained by performing modularity optimization in projected networks. Specifically, we show that the use of the concept of statistically validated network \cite{Tumminello2011} is useful to reveal subsets of nodes that defines cores of partitions of projected networks  with a high degree of precision.
The cores of partitions are statistically well defined, highly informative, and robust to incompleteness and errors of the bipartite system.    

The paper is organized as follows. In Sect. \ref{bench} we briefly describe the community detection procedure and we describe the generation of an artificial benchmark network. Sect. \ref{SVN} discusses the concept of statistically validated network. In Sect. \ref{methods} we present the two main indicators
used to compare partitions.  Sect. \ref{resbench} presents the results obtained with an artificial benchmark network whereas Sect. \ref{realnet} presents the results obtained with two real networks. Sect. \ref{conclusion} concludes.    

\section{Artificial benchmark network}\label{bench}

In the present study we focus on the community detection of a weighted projected network obtained from a bipartite network. 
We consider a community detection algorithm based on the maximization of modularity quality function. Modularity \cite{Newman2004,Newman2004b} is defined as
\begin{equation}
Q=\frac{1}{2m}\sum_{ij} \left[ A_{ij}-\frac{w_i w_j}{2m} \right] \delta(c_i,c_j)
\end{equation}
where $A_{ij}$ is the weighted adjacency matrix, $w_i=\sum_{j}A_{ij}$ is the strength of node $i$,  $2m=\sum_{i,j} A_{ij}$, and $c_i$ indicates the membership of community $i$. The weights of the projected networks that we are using in the present study are sometime called simple weights. For a pair of nodes $i$ and $j$ of set A of a bipartite network they are defined as the number of common neighbors of set B\footnote{The characteristics of the most appropriate null model to be used in the modularity maximization of the weighted projected network has been discussed in \cite{Guimera2007}. We have verified that the correction proposed in \cite{Guimera2007} is not crucial in our investigations and therefore, for the sake of simplicity, we are using in the present paper the null model originally introduced for unipartite networks.}. 

\begin{figure}[t]                                          
   \centering
	\subfigure[\label{Bench}]{{\includegraphics[width=.6\linewidth]{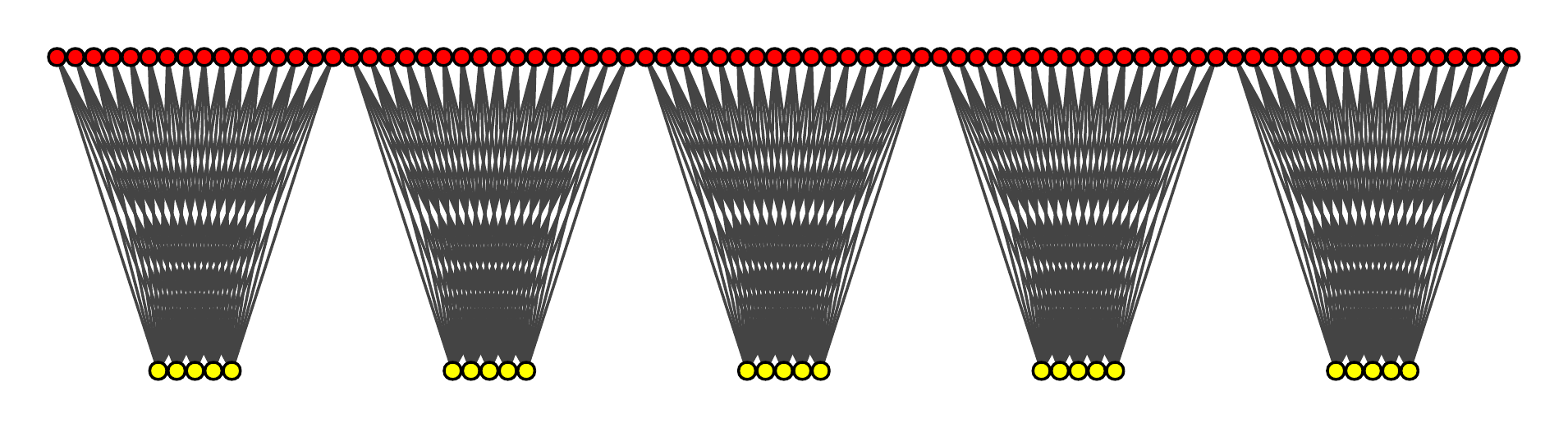} }}
	\qquad
    \subfigure[\label{OrigProj}]{{\includegraphics[width=.3\linewidth]{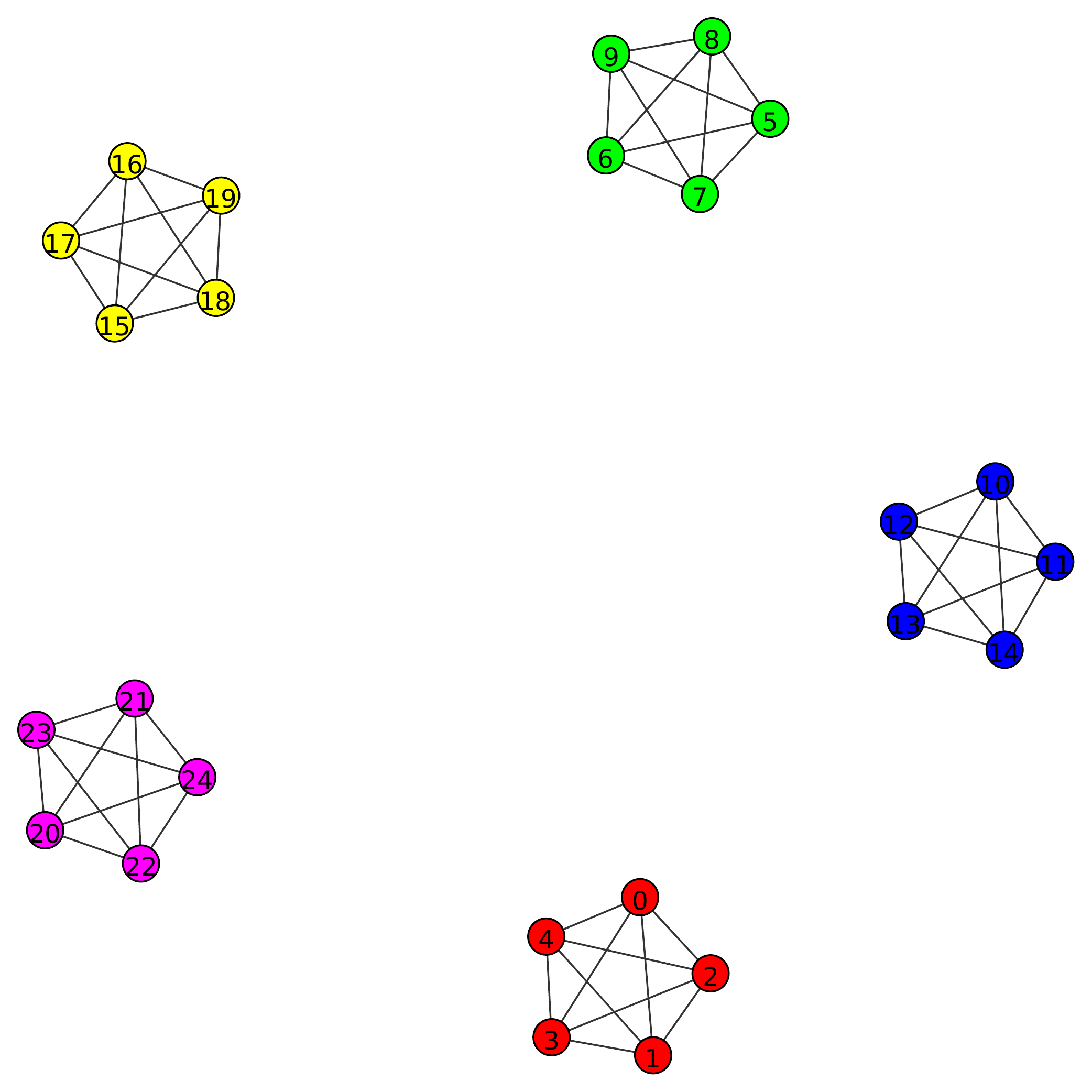}  }}
   \qquad
	\subfigure[\label{BenchNoise}]{{\includegraphics[width=.6\linewidth]{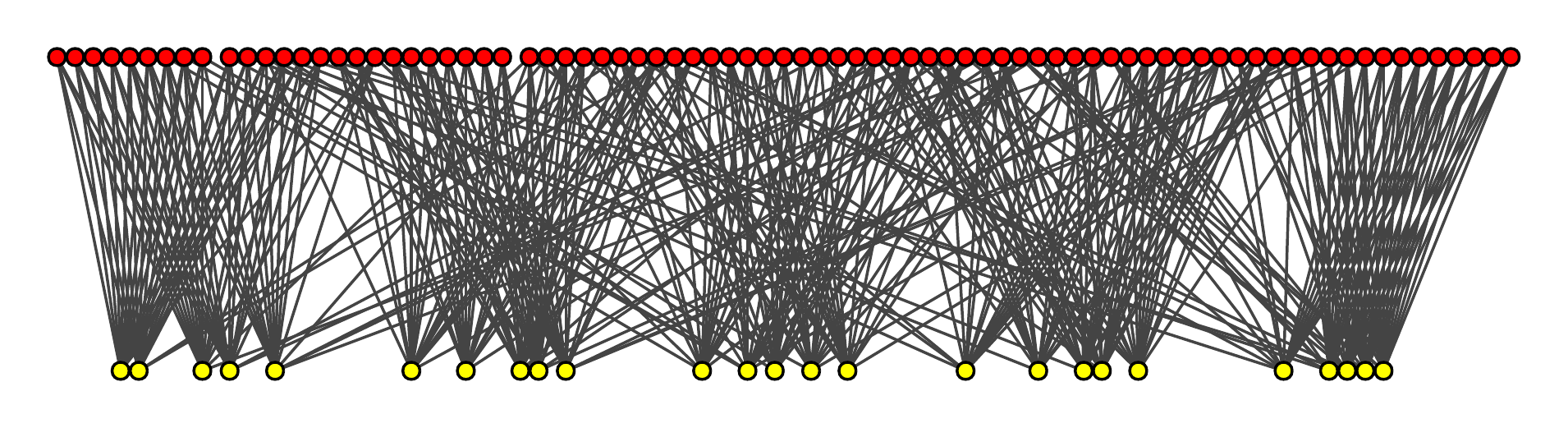}  }}
\caption{(a) Bipartite artificial benchmark network obtained with $q=5$, $S_A=5$, $S_B=16$, and $p_c=1$. Nodes in the bottom (top) row belongs to set A (B). (b) Network projection for the nodes of set A of the artificial benchmark of panel (a). (c) Bipartite artificial benchmark with $q=5$, $S_A=5$, $S_B=16$, $p_c=1$, and $p_r=0.2$.}	
	\label{fig:BO}
\end{figure}

We first illustrate our approach by considering an artificial benchmark network. Specifically, we generate a bipartite network with a well defined community structure as follows. Let $q$ be an integer defining the number of communities present in the artificial benchmark, and $\{s^A_1,\dots,s^A_q\}$ and $\{s^B_1,\dots,s^B_q\}$ be partitions of sets $A$ and $B$, respectively. In the present simulations the $q$ communities are all with the same number of nodes $A$ ($S_A$) and of nodes $B$ ($S_B$). Sets $A$ and $B$ have $qS_A$ and $qS_B$ nodes, respectively (see panel a) of Fig.~\ref{fig:BO}). 

We want to investigate the effect of missing or misclassified links in community detection. We therefore simulate artificial benchmark networks affected by missing or misclassified links to various degree. Specifically, for each bipartite clique of the network, our artificial benchmark network is obtained by connecting nodes of set $A$ to nodes of set $B$  with probability $p_c$, i.e. with a given probability of coverage of links ranging from 0 to 1.
The parameter $p_c$ therefore controls the degree of completeness of links present in the bipartite network. With this choice, the parameter $p_c$ also controls the density of the links of the bipartite network. This first procedure of the benchmark generation leads to $q$ disjoint bipartite components of the bipartite network (see panels a) and b) of Fig.~\ref{fig:BO} were we show an example of the artificial benchmark network generated with $q=5$, $S_A=5$, $S_B=16$, and $p_c=1$).

With the aim of modeling possible sources of randomness or errors present in datasets describing a real system, a second step in the generation of the artificial benchmark network is to randomize the bipartite network by using the following procedure. Let us call $p_r$ the probability that a link is misplaced due to some randomness or error.  For each node $i$ of set $A$ with $k_i$ links $p_r k_i$  links are on average selected and randomly linked to nodes of set B avoiding multiple links. The probability $p_r$ is therefore quantifying the uncertainty added to the generated artificial benchmark network. In the limit case when $p_r=0$ one gets back a network without errors. In the opposite limit of $p_r=1$ one obtains a completely random bipartite network which has no relationship with the underlying community structure. In panel c) of Fig.~\ref{fig:BO} we show an artificial benchmark network characterized by $q=5$, $S_A=5$, $S_B=16$, $p_c=1$, and $p_r=0.2$.

\section{Statistically validated networks}\label{SVN}

Several studies have recently selected a subset of links of a network on the basis of a statistical test considering a well defined null hypothesis \cite{Serrano2009,Tumminello2011,Hatzopoulos2015,Saracco2016,Gualdi2016}. These subsets have been called statistically validated networks \cite{Tumminello2011}. In this study, we filter the projected networks by using the approach of statistically validated networks introduced in \cite{Tumminello2011} and we use the filtered networks to select cores of communities present in the investigated network. Specifically, we perform a statistical test for each link of a projected network. A link between node $i$ and node $j$ is included in the projected statistically validated network when we reject a statistical test assuming a null hypothesis of random linking between node $i$ and node $j$  having a degree $k_i$ and $k_j$ in the original bipartite network respectively. Specifically, the null hypothesis is rejected if the weight of the link in the projected network, i.e. the number of common neighbors of nodes $i$ and $j$ of set A in the set B is higher and not statistically compatible with the expected value $k_i k_j /N_B$, where $k_i$ and $k_j$ are the degree of nodes $i$ and $j$ in the bipartite network and $N_B$ is the number of nodes of set B.

By mapping this problem into a urn problem it is possible to write down the probability of observing $x$  common neighbors of nodes $i$ and $j$ in set A under the null hypothesis of random connection preserving the heterogeneity of degree of nodes of set B. The probability of observing $x$ common neighbors between nodes $i$ and $j$ is given by the hypergeometric distribution
\begin{equation}
H(x|N_B,k_{i},k_{j}) = \frac{{k_{i}\choose x}{N_B -k_{i}\choose k_{j}-x}}{{N_B \choose k_{j}}}.
\label{HyperDist}
\end{equation}
Starting from this probability, it is possible to perform a  one-side statistical test and assign a $p$-value that determines the presence of a statistically validated link between a pair of nodes $i,j$ having $k_{ij}$ neighbors or more as
\begin{equation}
p_{i,j} =1 - \sum_{x=0}^{k_{ij}-1} H(x|N_B,k_{i},k_{j}).
\label{HyperTest}
\end{equation}

By performing the statistical test on all pairs of nodes of the projected network we are doing a multiple hypothesis test comparison.   Multiple hypothesis test comparisons need a multiple hypothesis test correction to control the level of false positives. The most restrictive multiple hypothesis test correction is the Bonferroni correction \cite{Hochberg1987} performed by setting the statistical threshold as  $\alpha_B = \alpha/N_t = 0.01/N_t$, where $\alpha$ is the chosen univariate threshold (in our case 0.01) and $N_t=N_A(N_A-1)/2$ where $N_A$ is the number of nodes of set A.

The Bonferroni correction minimizes the number of false positive but often does not guarantee sufficient accuracy (usually it provides a large number of false negative). The procedure controlling the false discovery rate (FDR)  \cite{Benjamini1995} reduces the number of false negative by controlling the expected proportion of rejected null hypothesis without significantly expanding the number of false positive. The  control of the FDR is realized as follows: p-values from all the $N_t$ tests are first arranged in increasing order ($p_1<p_2<...<p_k<...<p_{N_t}$). Starting from the the highest p-value one controls  the inequality $p_i \le i ~ \alpha_B$. If this inequality is first verified for a value $k^*$ all tests characterized by $k \le k_*$ are rejected. In the present study we use both the Bonferroni correction and the FDR correction.

\section{Comparing different partitions}\label{methods}

In the following sections we compare pair of partitions of linked nodes of a projected network.
We use for our comparison two widely used indicators. The first is the adjusted Rand index and the second is an adjusted version of a Wallace index. In other words, the comparison is done by considering adjusted versions of the accuracy and precision of the detection of pairs of nodes in a given partition compared with a reference partition.  In our comparison the number of true positive (TP) pairs is the number of pairs of nodes being in the same community both in the considered and in the reference partition. The number of false positive (FP) pairs is the  number of pairs of nodes being in the same community in the considered partition but in different communities in the reference partition. A pair of nodes is classified as true negative (TN) pair when each node of the pair does not belong to the same community both in the considered partition and in the reference partition. Lastly, a pair of nodes is classified as false negative (FN) pair when each node of the pair does not belong to the same community in the considered partition whereas both nodes belong to the same community in the reference partition. 

The Rand index \cite{Rand1971} is essentially the accuracy of the pair classification and it is defined as 
\begin{equation}
R=\frac{TP+TN}{TP+FP+TN+FN}
\end{equation}
The RAND index varies between zero (absence of any accuracy in the considered partition) and one (total accuracy in the partitioning). However, also in the presence of random partitioning a certain degree of accuracy can be obtained by chance. To take into account this possibility and adjusted version of the RAND index has been introduced \cite{Hubert1985}. The adjusted Rand index (ARI) is defined as
\begin{equation}
ARI=\frac{TP+TN-E\left[TP+TN\right]}{TP+FP+TN+FN-E\left[TP+TN\right]}
\end{equation}
where $E\left[TP+TN\right]$ is the expected value of the true pair classifications estimated between a random partition and the reference partition. For a random partition compared with another partition the value of the ARI is on average close to zero. Negative values of the index describe cases where the membership of the two partitions is more different than in a random case. 

By considering a set $N$ elements, and two partitions of these elements $X=\{X_{1},X_{2},\ldots ,X_{r}\}$ and $Y=\{Y_{1},Y_{2},\ldots ,Y_{s}\}$. By defining $n_{ij}$ as the number of elements in common between partition $X_{i}$ and $Y_{j}$, the ARI can also be written as
\begin{equation}
ARI=\frac{\sum_{i,j}{n_{ij}\choose 2} - \left[ \sum_{i}{a_{i}\choose 2} \sum_{j}{b_{j}\choose 2} \right] / {N\choose 2}}{ \frac{1}{2} \left[ \sum_{i}{a_{i}\choose 2}+ \sum_{j}{b_{j}\choose 2} \right]- \left[ \sum_{i}{a_{i}\choose 2} \sum_{j}{b_{j}\choose 2} \right] / {N\choose 2}}
\end{equation}
where $a_i=\sum_j^s n_{ij}$ and $b_j=\sum_i^r n_{ij}$.

\begin{figure}[!t]
   \centering
	\subfigure[\label{Nested}]{{\includegraphics[width=.40\linewidth]{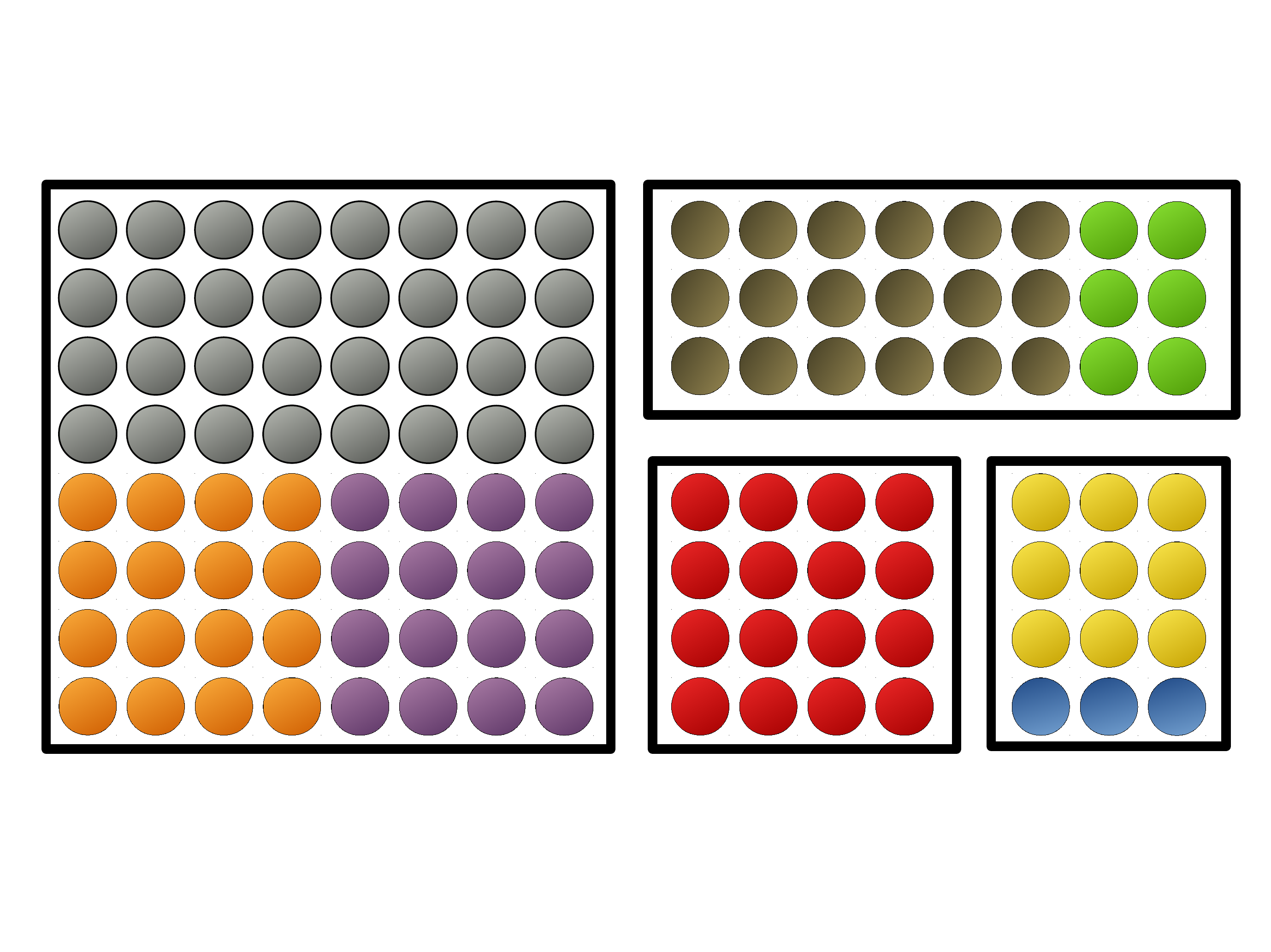} }}
	\qquad
    \subfigure[\label{NotNested1}]{{\includegraphics[width=.40\linewidth]{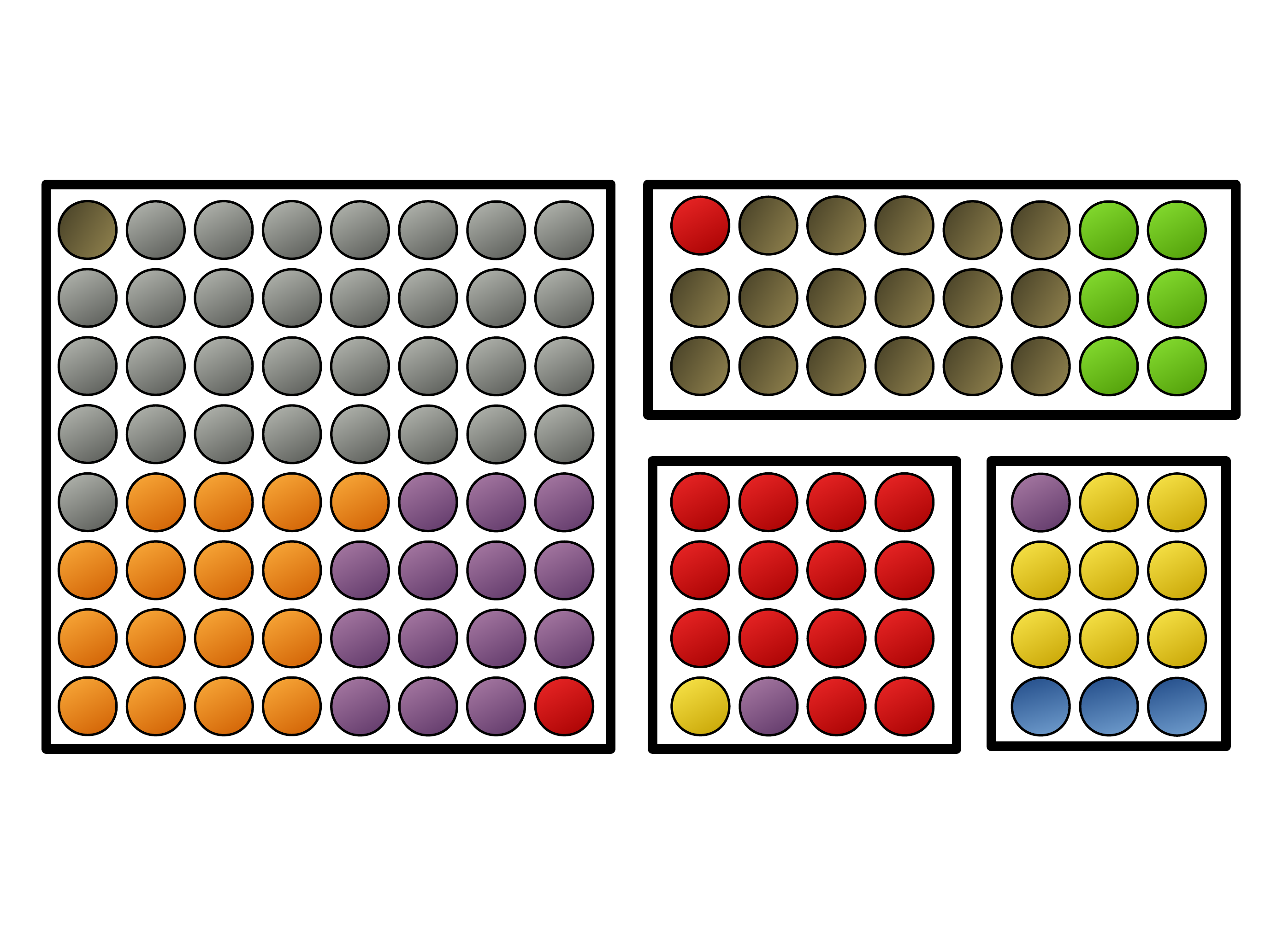} }}
    \qquad
	\subfigure[\label{NotNested2}]{{\includegraphics[width=.40\linewidth]{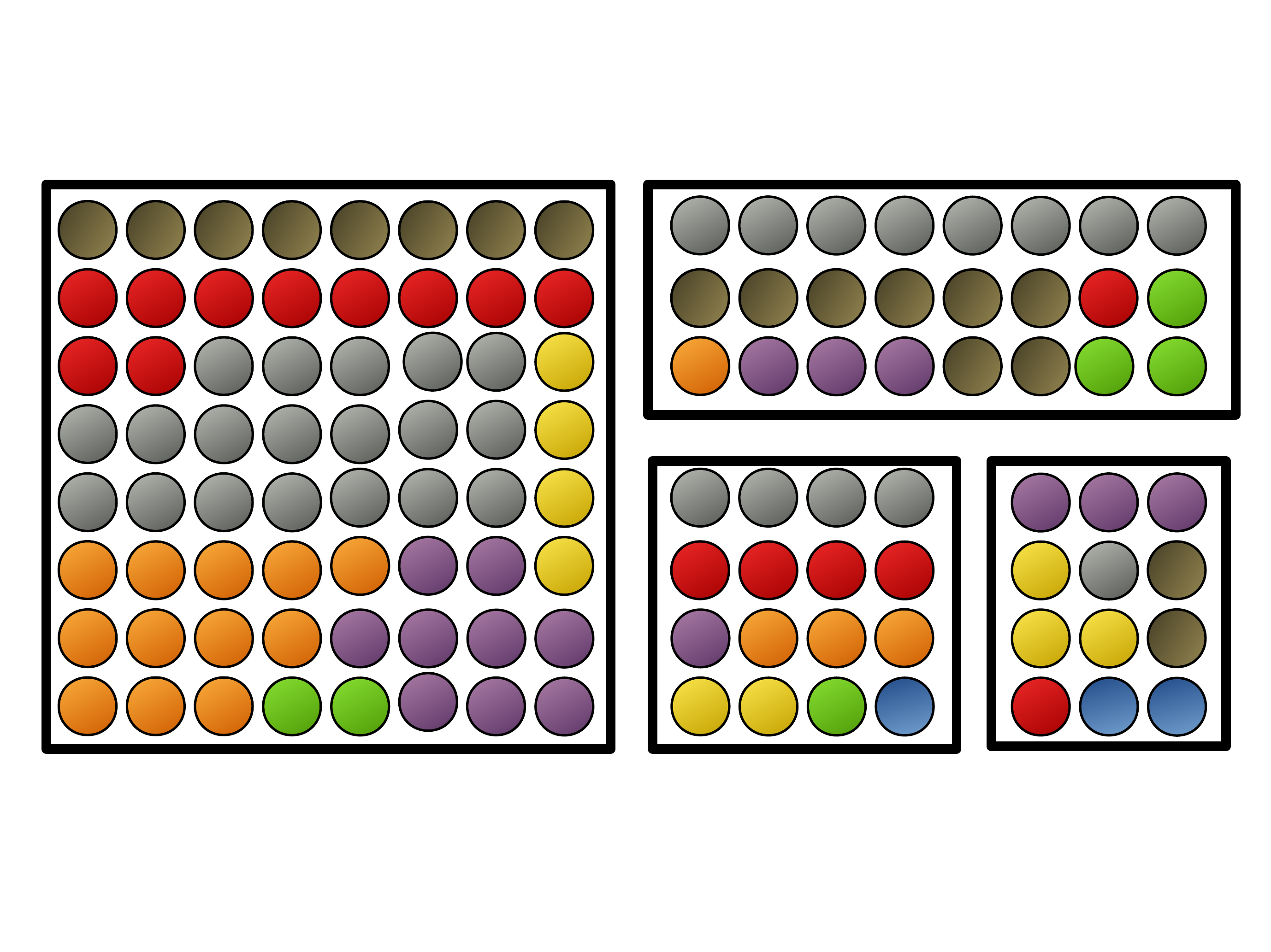}}}
	\caption{Three examples of comparison of a considered partition (membership of nodes indicated by different colors) with a reference partition (membership of nodes indicated by their position in different boxes). In the example a system of 116 nodes has 4 communities of different size in the reference partition (see four boxes with 64, 24, 16, and 12 nodes) and 8 communities of different size in the considered partition. This second partition is indicated by the colors of nodes. We have light gray (32 nodes), gray (18), orange (16), purple (16), red (16), yellow (9), green (6), and blue (3) groups. In the three examples the AWI assumes the values:  (a) $AWI = 1.0$, (b) $AWI = 0.88$, and (c) $AWI = 0.03$ }
	\label{fig:nest}
\end{figure}

The precision of the pairwise classification is defined as
\begin{equation}
P=\frac{TP}{TP+FP}
\end{equation}
When two memberships are compared pairwise the precision is usually addressed as one of the Wallace indices \cite{Wallace1983,Carrico2006}. Also for the case of the Wallace index one can consider an adjusted version of it. Hereafter we provide the definition of an adjusted version of the Wallace index that we call adjusted Wallace index (AWI)
\begin{equation}
AWI=\frac{TP-E\left[TP\right]}{TP+FP-E\left[TP\right]}
\end{equation}
where 
\begin{equation}
E\left[TP\right]=\frac{(TP+FP)(TP+FN)}{TP+FP+TN+FN}.
\end{equation} 
It is worth noting that the AWI is varying between $-\infty$ and one. A high value of the AWI indicates a high precision in selecting pairs of nodes that are belonging to the same community as defined in the reference partition.  In Fig. \ref{fig:nest} we provide an illustrative example of the estimation of the index. The reference partition is shown by grouping the membership of the nodes in different boxes. Specifically, the system of 116 nodes has 4 communities of different size (64, 24, 16, and 12 in the example). In the figure, the colors of nodes indicate the membership of the considered partition to be compared with the reference one. The considered partition has 8 communities indicated by different colors. In panel a) of Fig.   \ref{fig:nest} communities of the considered partition (labeled with colors) have pairs of nodes that are always contained in communities of the reference partition (labeled with boxes) and therefore the AWI is equal to one. In panel b) the membership of pairs of nodes of communities (colors) of the considered partition are only partially contained in communities of the reference partition (boxes). For example the red nodes are primarily in the bottom left box but two of them are with the largest and the second largest community in the reference partition respectively. In this second example the AWI is equal to 0.88 indicating a high but not perfect precision of the membership of pairs of nodes in the considered partition.  In panel c) the considered partition (colors) is quite different from the reference partition (boxes) and almost all boxes contain nodes of all colors. In this last case the AWI is close to zero (AWI=0.03), i.e. the value of the adjusted Wallace index is close to the one expected under a random distribution of nodes in the considered partition (colors).

\section{Results on an artificial benchmark}\label{resbench} 
                                               
We investigate the artificial network benchmark described in Section \ref{bench} by performing community detection on a projected network of it. Specifically, the community detection is performed on three different networks all of them obtained starting from the same bipartite network. The first is the weighted projected network (we address this network as the {\it Full} network connoting with this name the fact that for this network we are considering all links obtained from the projection).
The second network is a statistically validated network obtained with the procedure described in Section \ref{SVN} when the multiple hypothesis test correction is the Bonferroni correction. We address this network as the Bonferroni network. The third one is the statistically validated network obtained with the control of the FDR correction. We address this third type of network the FDR network. The Bonferroni network is a subgraph of the FDR network that is a subgraph of the Full network. 

For all three networks we perform a community detection by using modularity optimization. Specifically we  use the Louvain algorithm \cite{Blondel2008} and we analyze the partition associated with the highest value of modularity. 
It is worth noting that the role of the community detection algorithm is different for the Full network and for the SVNs. This is due to the fact that SVNs take the form of a large number of disconnected components and therefore for these networks the community detection algorithm is effective only on the largest of them.   

\begin{figure}[!t]
 	\hspace{0mm}%
   \centering
	\subfigure[\label{ARIq50}]{{\includegraphics[width=.850\linewidth]{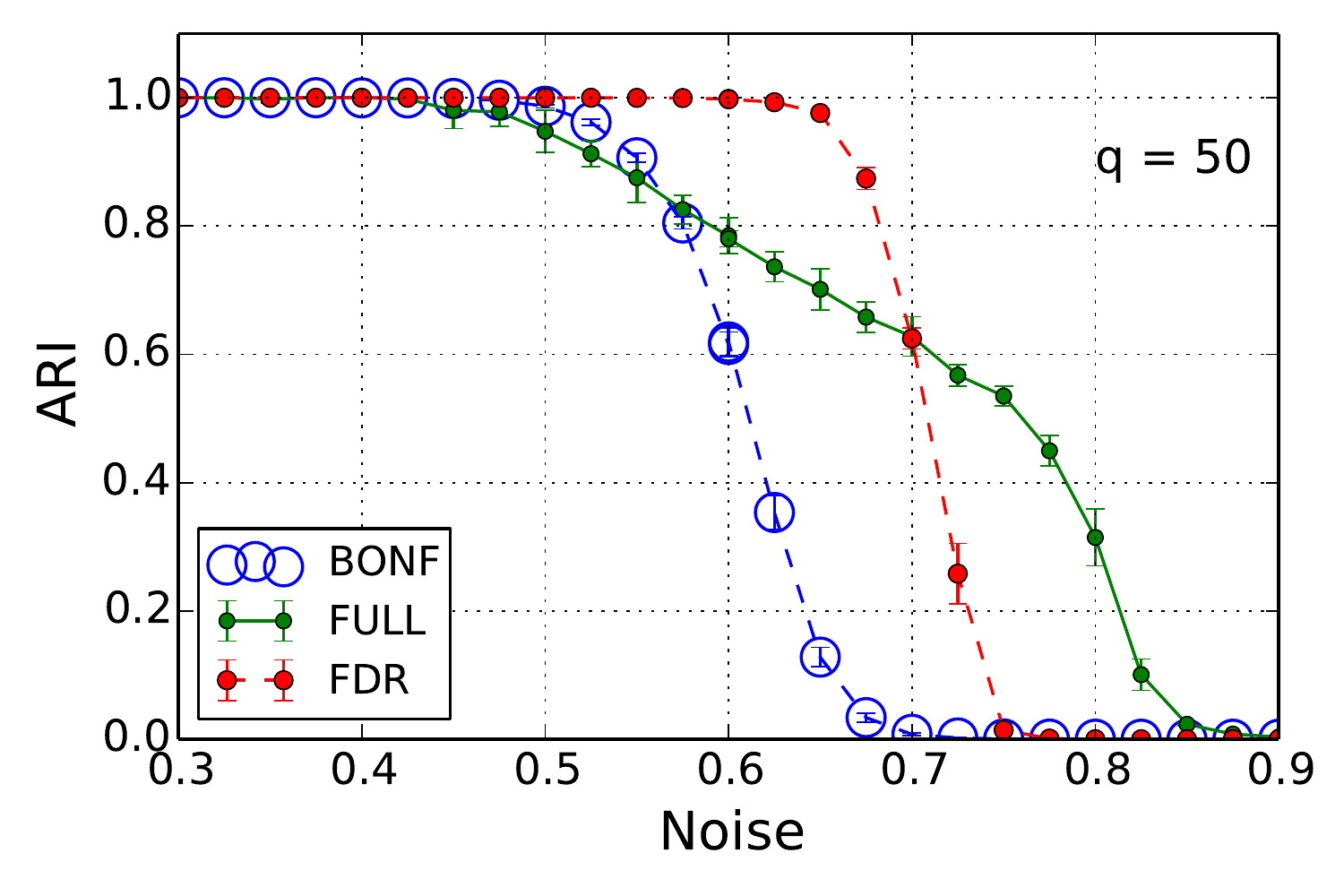} }}

	\subfigure[\label{AWIq20}]{{\includegraphics[width=.850\linewidth]{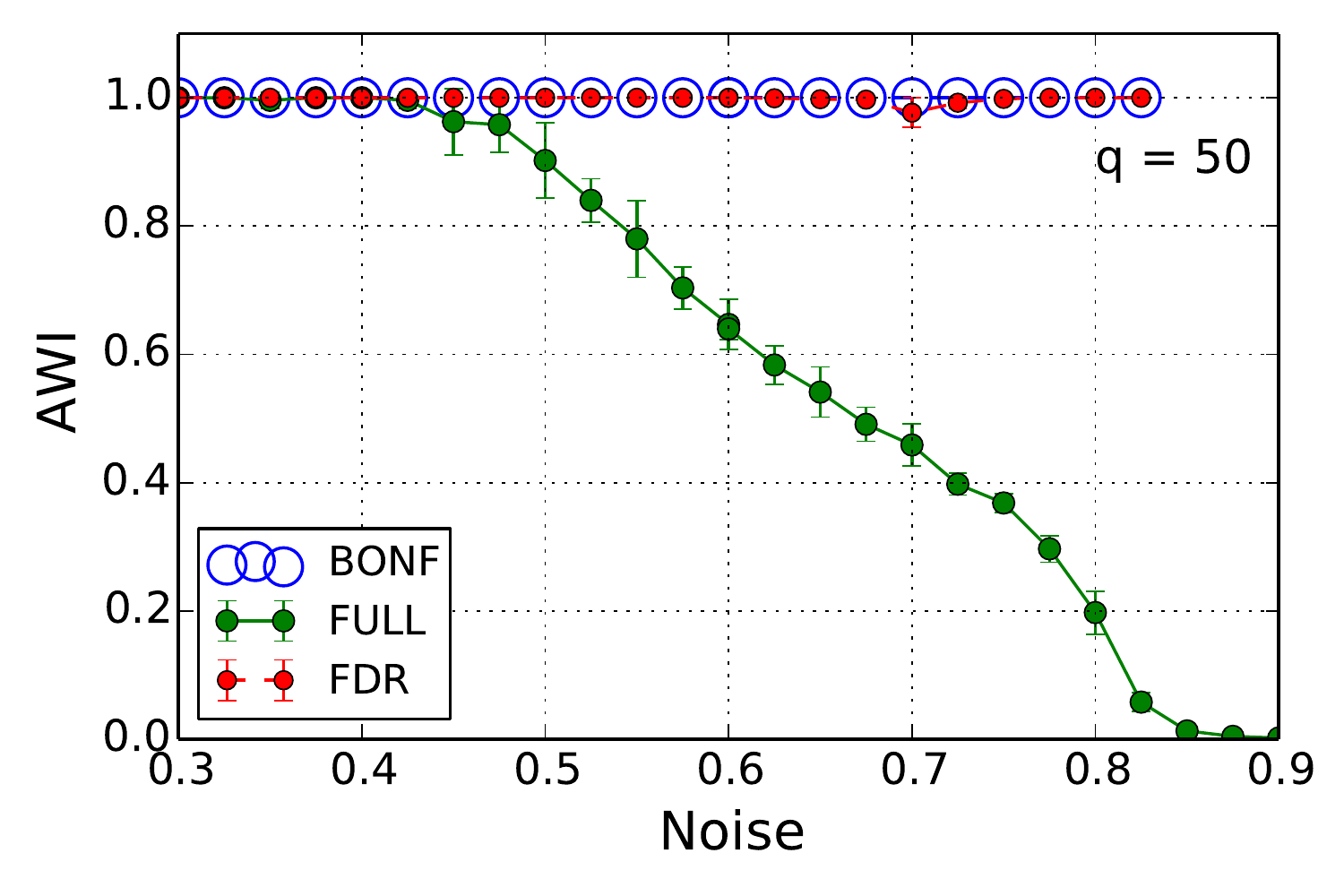} }}
	 \hspace{0mm}%
	\caption{ARI and AWI measured between the partition obtained by performing community detection for the three types of projected networks (i) Full (green symbols), (ii) FDR (red symbols) and (iii) Bonferroni (blue symbols) and the reference partition of the artificial benchmark. The benchmark has parameter $S_A=50$, $S_B=50$, $p_c=0.8$, $q=50$. Simulations and community detection are performed  for several values of $p_r$ ranging from 0.3 to 0.9 in steps of 0.025. Average value and one standard deviation error bar are obtained by performing the analysis on 10 different realizations.}
	\label{FigA}
\end{figure}

To take into account the stochastic nature of the algorithm and to verify the reproducibility of the obtained results we apply several times the algorithm by using a different initializing node sequence. With this approach the output of the Louvain algorithm is stochastic and different partitions can be obtained for different runs of the algorithm. In Fig. \ref{FigA} we show the ARI and AWI measured between the partition obtained by performing community detection of the three type of projected networks and the reference partition. Different versions of the benchmark were obtained by setting $S_A=50$, $S_B=50$, $p_c=0.8$, $q=50$ and several values of $p_r$ ranging from 0.3 to 0.9 in steps of 0.025. In the top panel of Fig. \ref{FigA} we show the ARI  as a function of the probability of misplacement $p_r$ of a link in the bipartite network. For the full network (green symbols) the ARI is close to one for low values of $p_r$ and starts to decreases for values of $p_r$ greater than 0.4. The ARI reaches values close to zero when $p_r$ is greater than 0.9. The misclassification of the community detection procedure is due to the fact that  the algorithm is not able to detect all communities of the reference partition due to the random rearrangement of links. Specifically, for high values of $p_r$ the errors done by the community detection algorithm concern the merging of some communities of the reference partition.

A similar pattern of success is observed for the partitions obtained with SVNs. In fact, for the FDR network (red symbols) we observe a value of the ARI close to one for low values of $p_r$ and close to zero for high values of it. It is worth noting that for the specific parameters of the benchmark there is an interval of $p_r$ ($0.5 \le p \le 0.7$) where the ARI of the FDR network is higher than the corresponding ARI value of the Full network. The Bonferroni network has an analogous pattern but the decrease of the ARI is observed for smaller values of $p_r$ ($p_r \approx 0.5$). It is worth noting that the reason of the decrease of the ARI for the FDR and the Bonferroni network is completely different from the one of the Full network. In fact for the partitions of these SVNs the ARI decreases because the statistical test loses power, the number of 
links decreases, and the number of isolated nodes increases  as a function of $p_r$. This implies that the number of disconnected subgraphs (present in the SVNs and/or detected by the Louvain algorithm) increases while the number of nodes connected decreases.     
 
The bottom panel of Fig. \ref{FigA} shows the AWI for the three types of networks. For the Full network, the pattern of the AWI is similar to the pattern of the ARI. It starts very close to one and decreases to zero starting from $p_r \approx 0.4$. The behavior of the AWI of the SVNs is quite different
supporting our previous conclusion that the reasons underlying the ARI behavior observed for the SVNs are different from the ones of the Full network. 
In fact AWI remains very close to 1 for high values of $p_r$ until abruptly reach zero when the SVNs becomes empty networks, i.e. all the nodes are isolated. In other words the precision of classification of pairs of nodes is always high for SVNs and the problem they have in providing informative partitions for high values of $p_r$ is not precision but rather accuracy. All the partitions provided by applying community detection to SVNs are statistically precise but the level of accuracy progressively decreases in the presence of high levels of link misplacements.

\begin{figure}[!t]
 	\hspace{0mm}%
   \centering
	\subfigure[\label{ARIq50}]{{\includegraphics[width=.850\linewidth]{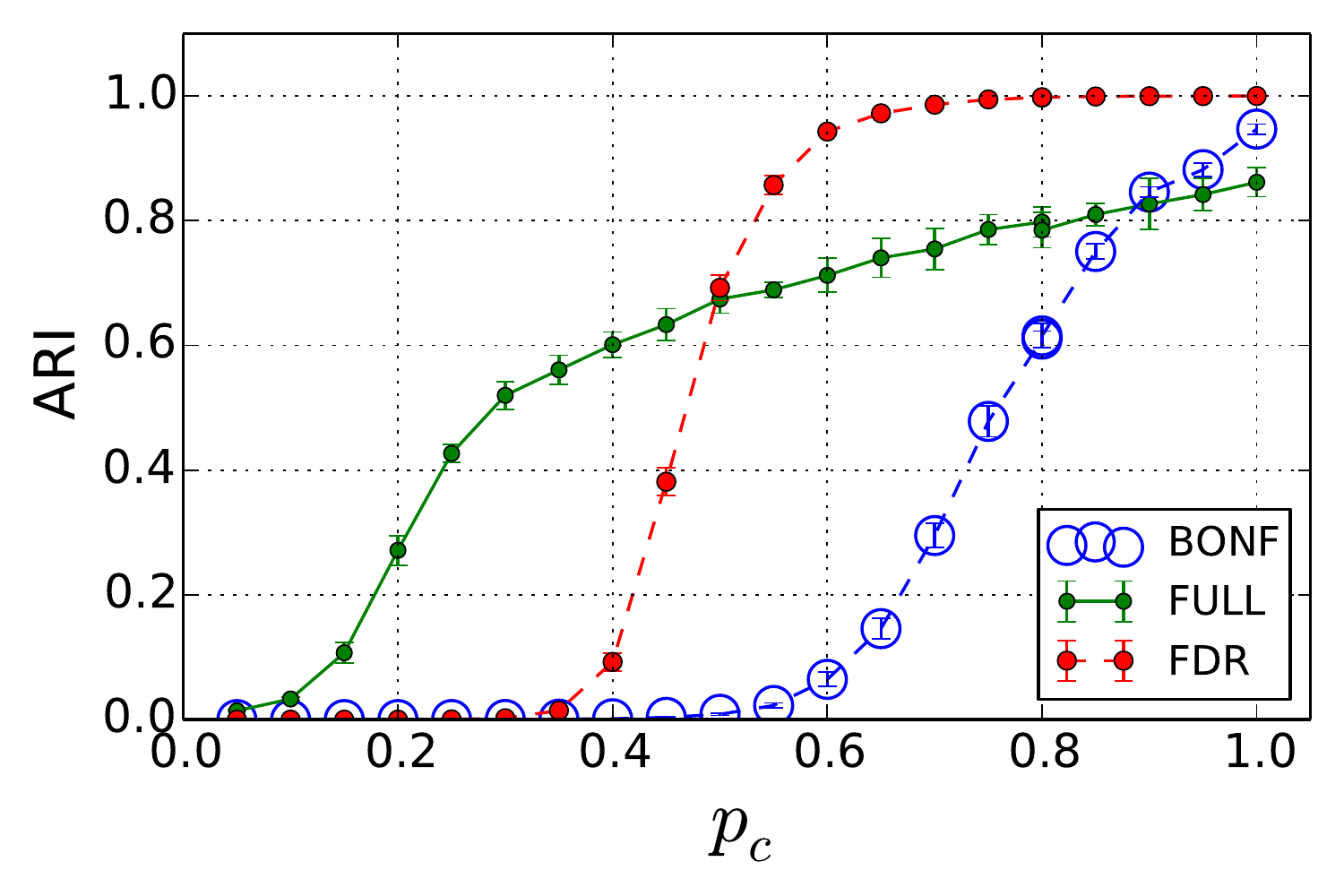} }}

	\subfigure[\label{AWIq20}]{{\includegraphics[width=.850\linewidth]{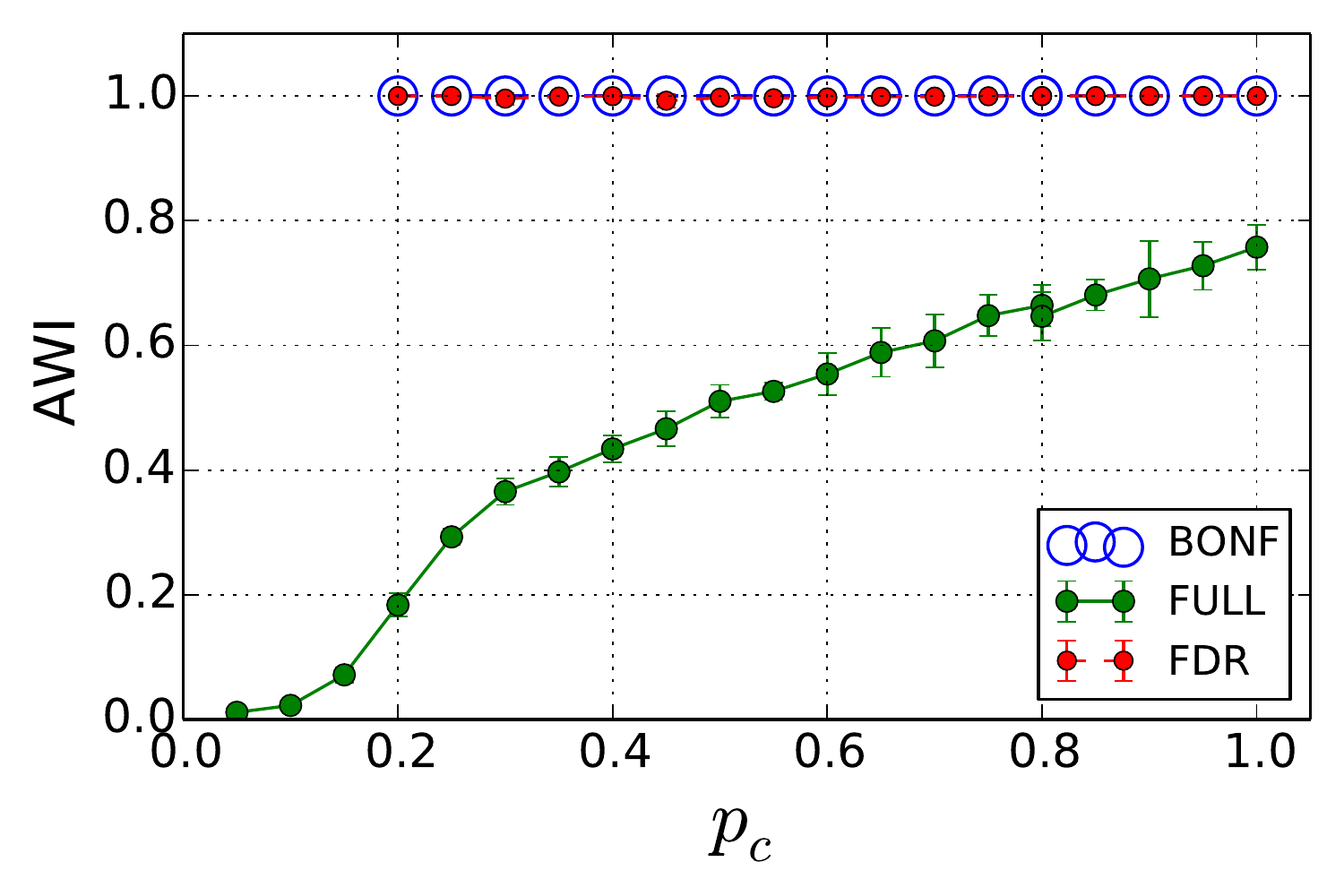} }}
	 \hspace{0mm}%
	\caption{ARI and AWI measured between the partition obtained by performing community detection for the three types of projected networks (i) Full (green symbols), (ii) FDR (red symbols) and (iii) Bonferroni (blue symbols) and the reference partition of the artificial benchmark. The benchmark has parameter $S_A=50$, $S_B=50$, $p_r=0.6$, $q=50$. Simulations and community detection are performed  for several values of $p_c$ ranging from 0 to 1.0 in steps of 0.025. Average value and one standard deviation error bar are obtained by performing the analysis on 10 different realizations.}
	\label{FigB}
\end{figure}

So far we have investigated the role of the link misplacement in the detection of communities of the artificial benchmark. Another cause of difficulty in community detection in real system can originate by insufficient coverage of the data. For this reason we have evaluated the performance of our approach for artificial benchmarks characterized by a different level of link coverage.    
In Fig. \ref{FigB} we show the ARI and AWI for simulations obtained by setting by setting $S_A=50$, $S_B=50$, $q=50$, $p_r=0.6$,  and for different values of $p_c$ ranging from 0 to 1 in steps of 0.05.

Panel (a) of Fig. \ref{FigB} shows that  the ability of the community detection algorithm to correctly detect reference communities of the benchmark decreases by decreasing $p_c$ both for the Full network and also for the SVNs. However also in this case the reason for this failure is different for the two approaches. In the case of the Full network the algorithm fails to detect the correct partition because it progressively merges several communities when $p_c$ decreases. On the other hand, the major problem observed for the partitions  obtained from SVNs is due to the fact that accuracy of the statistical validation decreases for values of $p_c$ lower than 0.7. In fact panel (b) of Fig. \ref{FigB} shows that for SVNs the AWI is always very close to one and therefore the failure is not due to a problem of precision but rather of accuracy as previously observed in the investigations of the artificial benchmark network performed as a function of $p_r$.

In summary, both as a function of $p_r$ and as a function of $p_c$ the partitions observed with the approach of SVNs are partitions which are very precise in classifying the membership of pairs of nodes although they might present a poor accuracy in the presence of high values of $p_r$ or low values of $p_c$. The membership  obtained by investigating the SVNs  can therefore be seen as statistically validated cores of the communities present in a given network.

A software generating artificial benchmark networks, calculating statistically validated projected networks in bipartite systems, and estimating AWI is accessible at the web page \cite{Bongiorno2017}.

\section{Real networks} \label{realnet}
We also investigate two widely studied real bipartite networks. The first is the bipartite network of authors and papers obtained analyzing the cond-mat archive  \cite{Newman2001}.  The second is the classic bipartite network of actors and movies obtained by using information present in the international movie data base (IMDB). 

\subsubsection{Co-authorship network}

We first investigate the co-authorship bipartite network. This bipartite network was constructed by Mark Newman by considering preprints posted in the condensed matter section of the arXiv E-Print Archive between 1995 and 1999. The dataset is available at the web page {\url{https://toreopsahl.com/datasets/}} and it consists of $16726$ authors and $22015$ papers. Our analysis is limited to the largest connected component of $13861$ authors and $19466$ papers. We project the bipartite network to obtain the projected network of authors. We also estimate the FDR SVN of authors. The Full network has $44619$ links and the FDR network has $7768$ links. We perform on them community detection with the Louvain algorithm. For each network, the community detection is performed by applying the algorithm $1000$ times with different initial conditions.

The $1000$ partitions obtained for the Full network have modularity ranging from $0.864$ to $0.867$. To investigate the degree of similarity among partitions of top values of modularity we select partitions with modularity higher then the one of the $99$ percentile of the $1000$ best outputs of Louvain algorithm. Specifically, we select  $10$ out $1000$ partitions of highest modularity.
We then estimate the ARI between all distinct pairs of these 10 partitions. These 45 pairs have an average mutual ARI of $0.65$ with values ranging between the value of $0.59$ (minimum) and $0.71$ (maximum).
As already noted in different investigations \cite{Good2010,Zhang2014} these partitions are quite different the one from the other in spite of the fact that the modularity of partitions is almost identical (bounded within the interval $0.8666,0.8670$). We obtain a quite different result when we consider the top 10 partitions obtained by performing community detection in the FDR SVN. In fact these 10 partitions are the same and the ARI among all of them is just one. It worth noting that the FDR partition is not fully contained in any partition obtained from the Full network. In fact the interval of the AWI index of the FDR with respect to the Full partition is quite far from one and it is covering a relatively limited interval of values ($0.57,0.66$). 

\begin{figure}[!t]
 	\hspace{0mm}%
   \centering
	\subfigure[\label{FigAriNewmanNoise}]{\includegraphics[width=.850\linewidth]{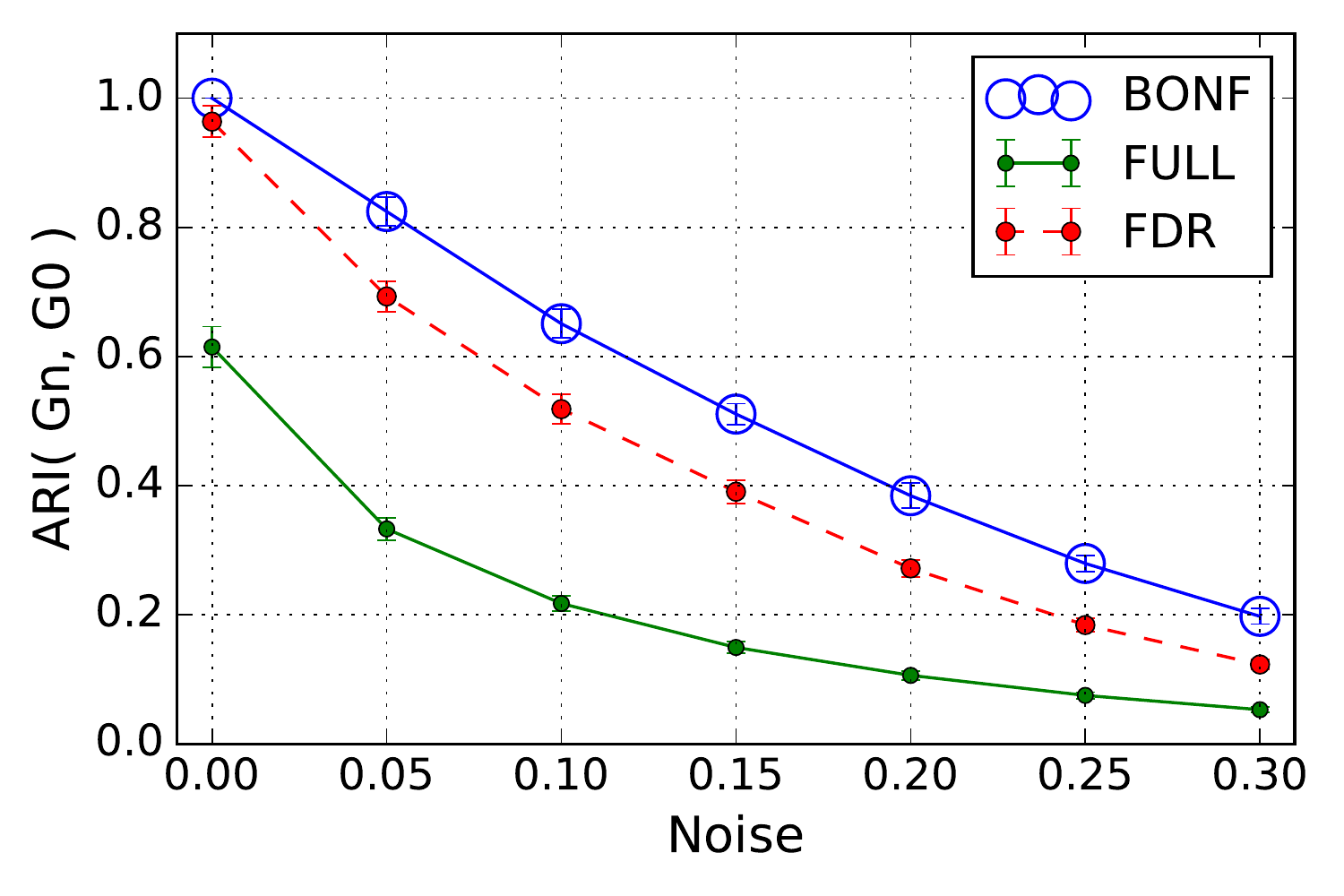} }
	\subfigure[\label{FigAwiNewmanNoise}]{\includegraphics[width=.850\linewidth]{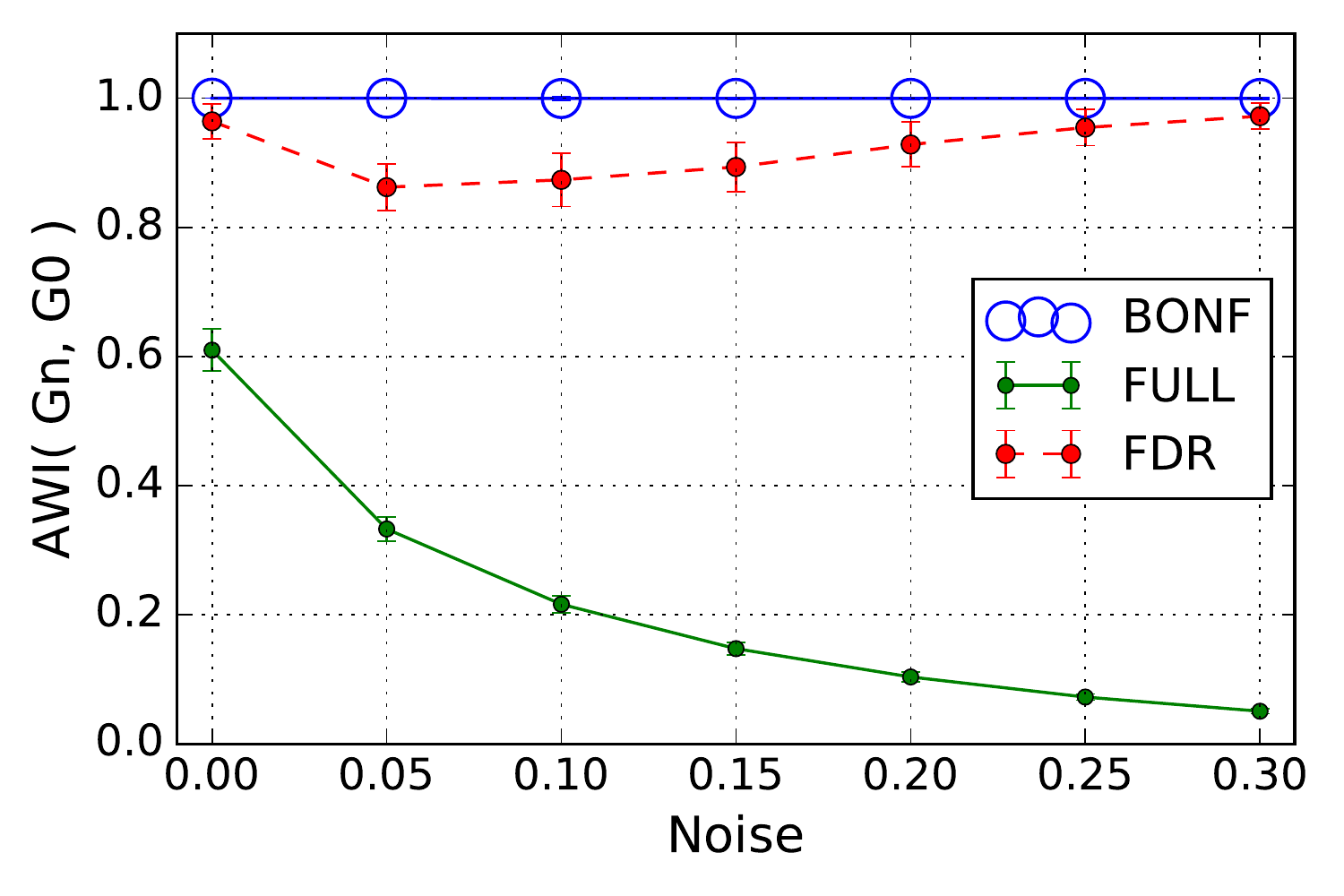} }
	\caption{Co-authorship database.  (a) Average ARI value between 100 partitions of the Full network (green symbols), the Bonferroni SVN (Blue symbols), and FDR SVN (red symbols) obtained as different stochastic realizations for each investigated value of $p_r$ and the best partition $G0$ obtained in the absence of additional noise. Error bar indicates one standard deviation. (b) Average AWI of the same partitions.}
	\label{FigNewmanNoise}
\end{figure}

By investigating the SVNs we therefore are able to extract {\it cores} of the communities that are statistically robust. These cores are also quite stable with respect to errors that might be present in the database. To make explicit this point we add some noise in the database by modifying it in a similar way to what we do with our artificial benchmark when we use values of $p_r$ different from zero. In panel (a) of Fig. \ref{FigNewmanNoise} we show the ARI between the best partition of the Full, that we label as G0, and 100 best partitions, that we label as Gn and that are obtained for each values of $p_r$ ranging from 0.05 to 0.3. In the same panel we also show the results of an analog investigation performed for the FDR SVN.  The partitions obtained from FDR SVNs are always significantly more robust to noise than the ones obtained by performing community detection in the Full network.     
In panel (b) of Fig. \ref{FigNewmanNoise} we show the AWI for the same numerical investigations. It is worth noting that the cores of communities detected by investigating the FDR SVN show a decrease of similarity (i.e. ARI values) with the uncorrupted partition G0 not due to decrease of precision but rather due to decrease of accuracy. In fact the AWI of FDR does not go below 0.85 for all values of $p_r$ whereas we observe values of the AWI as low as 0.1 of the partitions obtained from the Full network when $p_r=0.3$. In other words the informativeness of the detected cores of communities is robust with respect to noise added to the database. This behavior is similar to what we have observed for the artificial benchmark. 

\subsubsection{IMDB}

The second dataset we investigate is the classic bipartite system of actor and movies \cite{Watts1998}. We have  downloaded data about this system from the international movie data base (IMDB) ({\url{http://www.imdb.com/interfaces}}). From the information recorded in the database we obtain several bipartite networks. A link between an actor and a movies is considered if the actor played in that movie during a selected period of time. For our study we select all movies present in the database during the time period from 1950 to 2015, with the exception of  TV series, talk shows, animation, short and adult movies.

We perform our analyses for different periods of time defined by a time window of 5 years starting from 1950-1954. Within each selected time interval we construct a bipartite network considering movies released in that period and all actors that played in these movies. As for the previous system, our analysis is performed on the largest connected component observed in the considered period. The bipartite networks are projected into the movie side. The results of our investigations are summarized in Table \ref{Tab1}. Each row of the table refers to a different time period of investigation (see first column of the table). The size of the investigated projected networks changes over time from the lowest values of 9143 nodes and 686398 links to the highest values 127911 nodes and 1487598 links for the periods 1950-1954 and 2010-2014 respectively. The link density for the Full projected network of movies is ranging from $1.82~10^{-4}$  (for 2010-2014) to $1.64~10^{-2}$ (for 1950-1954), i.e. in all cases the projected networks are quite sparse. The Bonferroni and FDR SVNs are significantly more sparse than the Full network. In fact the percent of  SVNs links observed in Full network is always not exceeding 13.5 \% for FDR and 2.6 \% for Bonferroni (see the third and fourth column of Table \ref{Tab1}).  
 
 For each period of time and for the Full, the Bonferroni, and the FDR SVNs we have obtained 1000 output partitions by using the Louvain algorithm with different initial conditions. To evaluate the differences observed between pairs of partitions obtained we compute the ARI among the 10 partitions of the $99$ percentile of the $1000$ best outputs. The average value of the ARI is reported in the sixth, seventh, and eight  column of Table \ref{Tab1} for the Full, Bonferroni, and FDR networks respectively. The values of the ARI are always above 0.9 for all types of networks suggesting that for this database modularity optimization of the Full network is providing quite reliable results in most cases. In fact, values of the ARI lower than 0.97 are observed only for the last three time periods. The partitions obtained with the SVNs networks are
most stable than the partitions obtained from the Full network in most cases. Also in this case SVNs are detection cores of communities. This conclusion is also supported by the observed values of AWI between the Bonferroni and the Full network (nineth column of Table \ref{Tab1}), and between the FDR and the Full network (tenth column of Table \ref{Tab1}). In both cases the AWI is very close to one for all time periods except the last three ones when the modularity optimization of the Full becomes a bit less reliable.
 
\begin{figure}[!t]
 	\hspace{0mm}%
   \centering
	\subfigure[\label{FigAriIMDBNoise}]{\includegraphics[width=.850\linewidth]{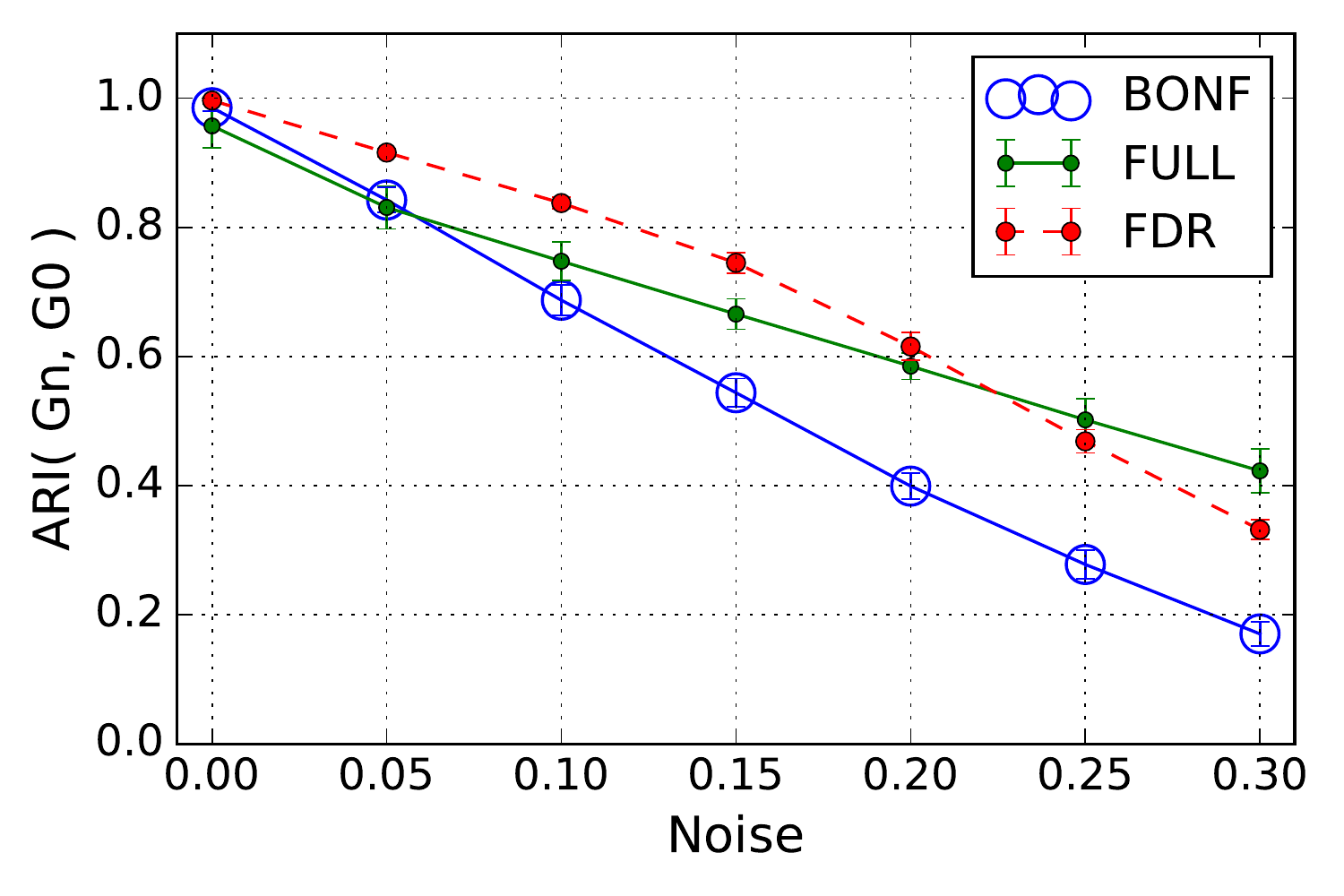} }
	\subfigure[\label{FigAwiIMDBNoise}]{\includegraphics[width=.8500\linewidth]{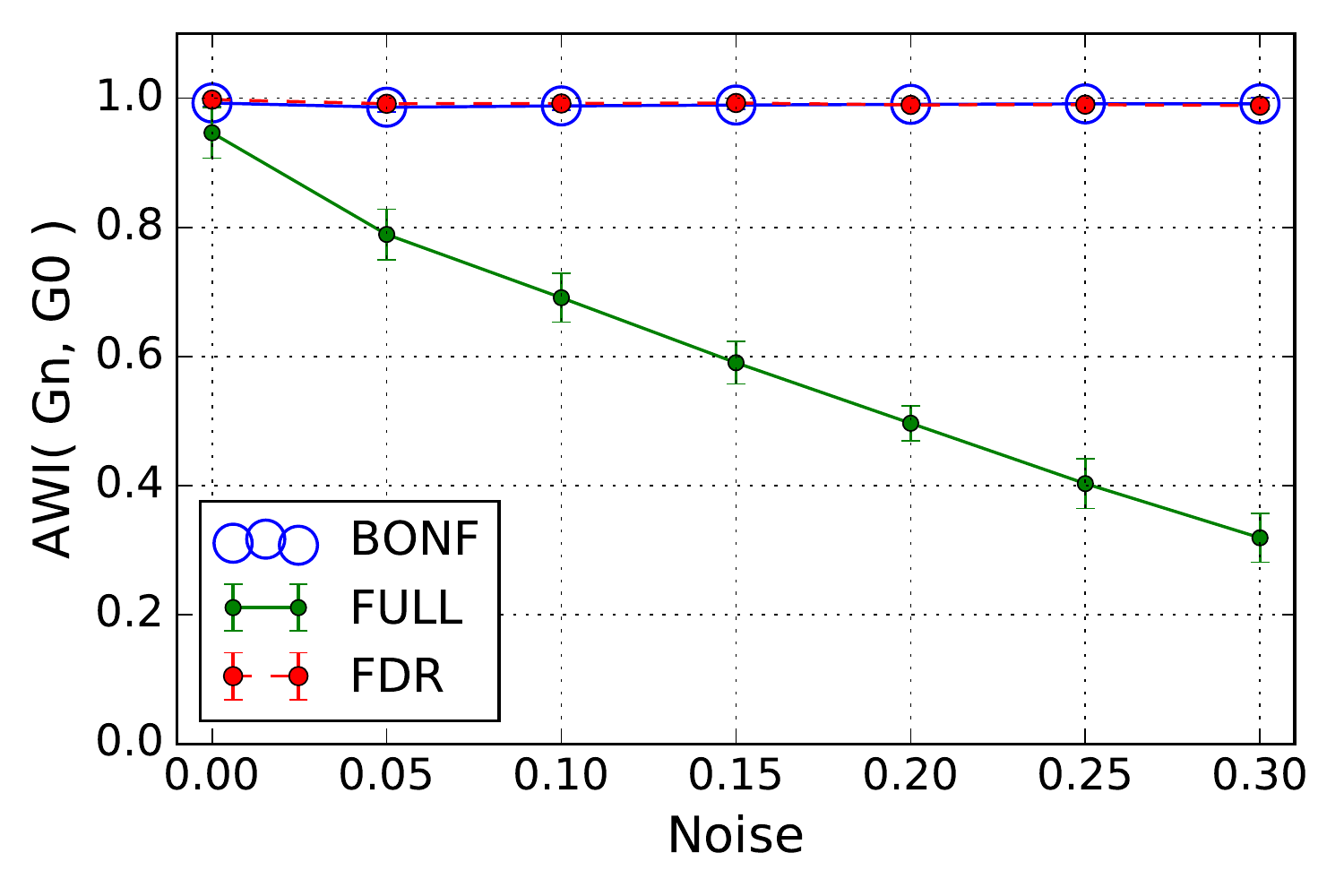} }
	\caption{IMDB database. Time period 1990-1994. (a) Average ARI value between 100 partitions of the Full network (green symbols), the Bonferroni SVN (Blue symbols), and FDR SVN (red symbols) obtained as different stochastic realizations for each investigated value of $p_r$ and the best partition $G0$ obtained in the absence of additional noise. Error bar indicates one standard deviation. (b) Average AWI of the same partitions.}
	\label{FigIMDBNoise}
\end{figure}

 Also for the IMDB bipartite networks of the period 1990-1994 we put additional noise in the bipartite network as we did with our artificial benchmark and with the co-authorship database.  In panel (a) of Fig. \ref{FigIMDBNoise} we show the average value of the ARI between 100 partitions of the Full obtained for values of $p_r$ ranging from 0.05 to 0.3 and the best partition $G0$ observed in the absence of noise. In the same panel we also show the results of an analog investigation performed for the Bonferroni and FDR SVNs.  The partitions obtained from FDR SVNs are for a large interval of $p_r$ significantly more similar and therefore more robust to noise that the ones obtained by performing community detection in the Full network.     
In panel (b) of Fig. \ref{FigIMDBNoise} we show the AWI for the same investigations. Again the AWI is close to one for the partitions of the SVNs supporting once again the conclusion about the high degree of precision of the method in the detection of cores of communities. As for previous cases, by combining the two information we conclude that the decreasing values of the ARI with the uncorrupted partition $G0$ for the Bonferroni and the FDR SVNs is not due to a decrease of precision but rather it is due to a decrease of accuracy of the SVN method.

\begin{table*}
\centering
\caption{Summary of IMDB investigations.}
\label{Tab1}
\begin{tabular}{|r|cccc|ccc|cc|}
\hline
 Time &   Nodes &    Links &  Bonf  \% &  FDR  \% &   AVG(ARI) &   AVG(ARI) &   AVG(ARI) &  AWI &  AWI \\
 period &  ~ &  ~ & of links &  of links &  Full &  Bonf &  FDR & (Bonf,Full) & (FDR,Full) \\
\hline
 1950-54 &    9143 &   686398 &                   1.4 &                  8.2 &  0.996 (0.993, 1.0)&  0.993 (0.984, 0.999)&  0.980 (0.959, 0.994) &            1.00 &           0.98 \\
 1955-59 &   11253 &   519240 &                   1.8 &                  9.1 &   0.992 (0.984, 0.999) &      1.0 (1.0,1.0) &        1.0 (1.0,1.0) &            1.00 &           0.98 \\
 1960-64 &   12392 &   506639 &                   1.9 &                 10.7 &   0.998 (0.995, 1.0) &      1.0 (1.0,1.0) &   0.990 (0.978, 1.0) &            1.00 &           0.97 \\
 1965-69 &   14782 &   633135 &                   2.1 &                 10.7 &  0.978 (0.961, 0.995) &  1.0 (1.0,1.0) &    0.995 (0.987, 0.998)   &            1.00 &           0.98 \\
 1970-74 &   15958 &   620634 &                   2.2 &                 11.1 &  0.983 (0.964, 0.997) &  0.989 (0.979, 1.0) &   0.998 (0.995, 1.0)  &            1.00 &           0.97 \\
 1975-79 &   14996 &   522389 &                   2.6 &                 13.3 &  0.970 (0.920, 0.993) &  0.999 (0.997, 1.0) &  0.996 (0.989, 1.0)     &            0.99 &           0.95 \\
 1980-84 &   15401 &   485082 &                   2.5 &                 13.5 &  0.995 (0.992, 0.998) &      1.0 (1.0,1.0) &  0.995 (0.990, 1.0) &            1.00 &           0.95 \\
 1985-89 &   16846 &   569253 &                   2.1 &                 13.2 &  0.990 (0.984, 0.997) &      1.0 (1.0,1.0) &    0.984 (0.968, 0.999) &            1.00 &           0.93 \\
 1990-94 &   17001 &   458604 &                   1.9 &                 10.2 &  0.985 (0.975, 0.993) &  0.998 (0.997, 1.0) &  0.999 (0.997, 1.0)    &            0.99 &           0.98 \\
 1995-99 &   20311 &   402736 &                   1.4 &                  7.1 &  0.982 (0.973, 0.991) &      1.0 (1.0,1.0) &    1.0 (1.0,1.0) &            1.00 &           0.97 \\
 2000-04 &   31231 &   470828 &                   1.4 &                  7.2 &  0.966 (0.952, 0.979) &      1.0 (1.0,1.0) &    0.997 (0.993, 1.0) &            0.98 &           0.93 \\
 2005-09 &   62496 &   788713 &                   1.5 &                  5.7 &  0.952 (0.937, 0.967) &      1.0 (1.0,1.0) &    0.941 (0.905, 0.977) &            0.93 &           0.73 \\
 2010-14 &  127911 &  1487598 &                   1.1 &                  4.4 &  0.940 (0.912, 0.957) &  0.992 (0.984, 1.0) &  0.949 (0.919, 0.987)    &            0.88 &           0.71 \\
\hline
\end{tabular}
\end{table*}

\section{Conclusions} \label{conclusion}
We have shown that information present in a bipartite network can be used to detect cores of communities (i.e. clusters) of each set of the bipartite system. The detected cores are highly stable and the detection of them is highly precise although the methodology can, in same cases, be of low accuracy. The cores of communities are found by considering statistically validated networks obtained starting from the bipartite network. The information carried by these statistically validated network is therefore highly informative and could be used to detect membership of the investigated set that are robust with respect to the presence of errors or missing entries in the database. The usefulness of the statistical validation approach can be assessed by using a measure of similarity between pairs of partitions that are obtained by a stochastic community detection algorithm and that differ between them only for a tiny value of the quality function of the algorithm. Here we use the ARI. In the presence of partitions characterized by very similar values of the quality function and presenting low values of the ARI between them one should consider informative only those subsets of partitions that are statistically stable. We propose that in these cases investigators focus on cores of the partitions obtained by performing community detection on SVNs.  In the present study we have considered an algorithm based on modularity optimization but we believe that our results are general and not strictly related to the chosen algorithm. They should be valid for any algorithm based on the maximization of a quality function.

\newpage

{

\bibliographystyle{apsrev4-1} 
\bibliography{BLMM_bibtex.bib} 

\begin{thebibliography}{33}%
\makeatletter
\providecommand \@ifxundefined [1]{%
 \@ifx{#1\undefined}
}%
\providecommand \@ifnum [1]{%
 \ifnum #1\expandafter \@firstoftwo
 \else \expandafter \@secondoftwo
 \fi
}%
\providecommand \@ifx [1]{%
 \ifx #1\expandafter \@firstoftwo
 \else \expandafter \@secondoftwo
 \fi
}%
\providecommand \natexlab [1]{#1}%
\providecommand \enquote  [1]{``#1''}%
\providecommand \bibnamefont  [1]{#1}%
\providecommand \bibfnamefont [1]{#1}%
\providecommand \citenamefont [1]{#1}%
\providecommand \href@noop [0]{\@secondoftwo}%
\providecommand \href [0]{\begingroup \@sanitize@url \@href}%
\providecommand \@href[1]{\@@startlink{#1}\@@href}%
\providecommand \@@href[1]{\endgroup#1\@@endlink}%
\providecommand \@sanitize@url [0]{\catcode `\\12\catcode `\$12\catcode
  `\&12\catcode `\#12\catcode `\^12\catcode `\_12\catcode `\%12\relax}%
\providecommand \@@startlink[1]{}%
\providecommand \@@endlink[0]{}%
\providecommand \url  [0]{\begingroup\@sanitize@url \@url }%
\providecommand \@url [1]{\endgroup\@href {#1}{\urlprefix }}%
\providecommand \urlprefix  [0]{URL }%
\providecommand \Eprint [0]{\href }%
\providecommand \doibase [0]{http://dx.doi.org/}%
\providecommand \selectlanguage [0]{\@gobble}%
\providecommand \bibinfo  [0]{\@secondoftwo}%
\providecommand \bibfield  [0]{\@secondoftwo}%
\providecommand \translation [1]{[#1]}%
\providecommand \BibitemOpen [0]{}%
\providecommand \bibitemStop [0]{}%
\providecommand \bibitemNoStop [0]{.\EOS\space}%
\providecommand \EOS [0]{\spacefactor3000\relax}%
\providecommand \BibitemShut  [1]{\csname bibitem#1\endcsname}%
\let\auto@bib@innerbib\@empty
\bibitem [{\citenamefont {Fortunato}\ and\ \citenamefont
  {Hric}(2016)}]{Fortunato2016}%
  \BibitemOpen
  \bibfield  {author} {\bibinfo {author} {\bibfnamefont {S.}~\bibnamefont
  {Fortunato}}\ and\ \bibinfo {author} {\bibfnamefont {D.}~\bibnamefont
  {Hric}},\ }\href@noop {} {\bibfield  {journal} {\bibinfo  {journal} {Physics
  Reports}\ }\textbf {\bibinfo {volume} {659}},\ \bibinfo {pages} {1} (\bibinfo
  {year} {2016})}\BibitemShut {NoStop}%
\bibitem [{\citenamefont {Fortunato}(2010)}]{Fortunato2010}%
  \BibitemOpen
  \bibfield  {author} {\bibinfo {author} {\bibfnamefont {S.}~\bibnamefont
  {Fortunato}},\ }\href@noop {} {\bibfield  {journal} {\bibinfo  {journal}
  {Physics reports}\ }\textbf {\bibinfo {volume} {486}},\ \bibinfo {pages} {75}
  (\bibinfo {year} {2010})}\BibitemShut {NoStop}%
\bibitem [{\citenamefont {Newman}(2010)}]{Newman2010}%
  \BibitemOpen
  \bibfield  {author} {\bibinfo {author} {\bibfnamefont {M.}~\bibnamefont
  {Newman}},\ }\href@noop {} {\emph {\bibinfo {title} {Networks: an
  introduction}}}\ (\bibinfo  {publisher} {Oxford University Press, Oxford},\
  \bibinfo {year} {2010})\BibitemShut {NoStop}%
\bibitem [{\citenamefont {Barab\'asi}(2016)}]{Barabasi2016}%
  \BibitemOpen
  \bibfield  {author} {\bibinfo {author} {\bibfnamefont {A.}~\bibnamefont
  {Barab\'asi}},\ }\href@noop {} {\emph {\bibinfo {title} {Network Science}}}\
  (\bibinfo  {publisher} {Cambridge university press.},\ \bibinfo {year}
  {2016})\BibitemShut {NoStop}%
\bibitem [{\citenamefont {Newman}\ and\ \citenamefont
  {Girvan}(2004)}]{Newman2004}%
  \BibitemOpen
  \bibfield  {author} {\bibinfo {author} {\bibfnamefont {M.~E.}\ \bibnamefont
  {Newman}}\ and\ \bibinfo {author} {\bibfnamefont {M.}~\bibnamefont
  {Girvan}},\ }\href@noop {} {\bibfield  {journal} {\bibinfo  {journal}
  {Physical review E}\ }\textbf {\bibinfo {volume} {69}},\ \bibinfo {pages}
  {026113} (\bibinfo {year} {2004})}\BibitemShut {NoStop}%
\bibitem [{\citenamefont {Blondel}\ \emph {et~al.}(2008)\citenamefont
  {Blondel}, \citenamefont {Guillaume}, \citenamefont {Lambiotte},\ and\
  \citenamefont {Lefebvre}}]{Blondel2008}%
  \BibitemOpen
  \bibfield  {author} {\bibinfo {author} {\bibfnamefont {V.~D.}\ \bibnamefont
  {Blondel}}, \bibinfo {author} {\bibfnamefont {J.-L.}\ \bibnamefont
  {Guillaume}}, \bibinfo {author} {\bibfnamefont {R.}~\bibnamefont
  {Lambiotte}}, \ and\ \bibinfo {author} {\bibfnamefont {E.}~\bibnamefont
  {Lefebvre}},\ }\href@noop {} {\bibfield  {journal} {\bibinfo  {journal}
  {Journal of statistical mechanics: theory and experiment}\ }\textbf {\bibinfo
  {volume} {2008}},\ \bibinfo {pages} {P10008} (\bibinfo {year}
  {2008})}\BibitemShut {NoStop}%
\bibitem [{\citenamefont {Fortunato}\ and\ \citenamefont
  {Barthelemy}(2007)}]{Fortunato2007}%
  \BibitemOpen
  \bibfield  {author} {\bibinfo {author} {\bibfnamefont {S.}~\bibnamefont
  {Fortunato}}\ and\ \bibinfo {author} {\bibfnamefont {M.}~\bibnamefont
  {Barthelemy}},\ }\href@noop {} {\bibfield  {journal} {\bibinfo  {journal}
  {Proceedings of the National Academy of Sciences}\ }\textbf {\bibinfo
  {volume} {104}},\ \bibinfo {pages} {36} (\bibinfo {year} {2007})}\BibitemShut
  {NoStop}%
\bibitem [{\citenamefont {Reichardt}\ and\ \citenamefont
  {Bornholdt}(2006)}]{Reichardt2006}%
  \BibitemOpen
  \bibfield  {author} {\bibinfo {author} {\bibfnamefont {J.}~\bibnamefont
  {Reichardt}}\ and\ \bibinfo {author} {\bibfnamefont {S.}~\bibnamefont
  {Bornholdt}},\ }\href@noop {} {\bibfield  {journal} {\bibinfo  {journal}
  {Physical Review E}\ }\textbf {\bibinfo {volume} {74}},\ \bibinfo {pages}
  {016110} (\bibinfo {year} {2006})}\BibitemShut {NoStop}%
\bibitem [{\citenamefont {Arenas}\ \emph {et~al.}(2008)\citenamefont {Arenas},
  \citenamefont {Fernandez},\ and\ \citenamefont {Gomez}}]{Arenas2008}%
  \BibitemOpen
  \bibfield  {author} {\bibinfo {author} {\bibfnamefont {A.}~\bibnamefont
  {Arenas}}, \bibinfo {author} {\bibfnamefont {A.}~\bibnamefont {Fernandez}}, \
  and\ \bibinfo {author} {\bibfnamefont {S.}~\bibnamefont {Gomez}},\
  }\href@noop {} {\bibfield  {journal} {\bibinfo  {journal} {New Journal of
  Physics}\ }\textbf {\bibinfo {volume} {10}},\ \bibinfo {pages} {053039}
  (\bibinfo {year} {2008})}\BibitemShut {NoStop}%
\bibitem [{\citenamefont {Lancichinetti}\ and\ \citenamefont
  {Fortunato}(2011)}]{Lancichinetti2011a}%
  \BibitemOpen
  \bibfield  {author} {\bibinfo {author} {\bibfnamefont {A.}~\bibnamefont
  {Lancichinetti}}\ and\ \bibinfo {author} {\bibfnamefont {S.}~\bibnamefont
  {Fortunato}},\ }\href@noop {} {\bibfield  {journal} {\bibinfo  {journal}
  {Physical review E}\ }\textbf {\bibinfo {volume} {84}},\ \bibinfo {pages}
  {066122} (\bibinfo {year} {2011})}\BibitemShut {NoStop}%
\bibitem [{\citenamefont {Good}\ \emph {et~al.}(2010)\citenamefont {Good},
  \citenamefont {de~Montjoye},\ and\ \citenamefont {Clauset}}]{Good2010}%
  \BibitemOpen
  \bibfield  {author} {\bibinfo {author} {\bibfnamefont {B.~H.}\ \bibnamefont
  {Good}}, \bibinfo {author} {\bibfnamefont {Y.-A.}\ \bibnamefont
  {de~Montjoye}}, \ and\ \bibinfo {author} {\bibfnamefont {A.}~\bibnamefont
  {Clauset}},\ }\href@noop {} {\bibfield  {journal} {\bibinfo  {journal}
  {Physical Review E}\ }\textbf {\bibinfo {volume} {81}},\ \bibinfo {pages}
  {046106} (\bibinfo {year} {2010})}\BibitemShut {NoStop}%
\bibitem [{\citenamefont {Zhang}\ and\ \citenamefont
  {Moore}(2014)}]{Zhang2014}%
  \BibitemOpen
  \bibfield  {author} {\bibinfo {author} {\bibfnamefont {P.}~\bibnamefont
  {Zhang}}\ and\ \bibinfo {author} {\bibfnamefont {C.}~\bibnamefont {Moore}},\
  }\href@noop {} {\bibfield  {journal} {\bibinfo  {journal} {Proceedings of the
  National Academy of Sciences}\ }\textbf {\bibinfo {volume} {111}},\ \bibinfo
  {pages} {18144} (\bibinfo {year} {2014})}\BibitemShut {NoStop}%
\bibitem [{\citenamefont {Barber}(2007)}]{Barber2007}%
  \BibitemOpen
  \bibfield  {author} {\bibinfo {author} {\bibfnamefont {M.~J.}\ \bibnamefont
  {Barber}},\ }\href@noop {} {\bibfield  {journal} {\bibinfo  {journal}
  {Physical Review E}\ }\textbf {\bibinfo {volume} {76}},\ \bibinfo {pages}
  {066102} (\bibinfo {year} {2007})}\BibitemShut {NoStop}%
\bibitem [{\citenamefont {Larremore}\ \emph {et~al.}(2014)\citenamefont
  {Larremore}, \citenamefont {Clauset},\ and\ \citenamefont
  {Jacobs}}]{Larremore2014}%
  \BibitemOpen
  \bibfield  {author} {\bibinfo {author} {\bibfnamefont {D.~B.}\ \bibnamefont
  {Larremore}}, \bibinfo {author} {\bibfnamefont {A.}~\bibnamefont {Clauset}},
  \ and\ \bibinfo {author} {\bibfnamefont {A.~Z.}\ \bibnamefont {Jacobs}},\
  }\href@noop {} {\bibfield  {journal} {\bibinfo  {journal} {Physical Review
  E}\ }\textbf {\bibinfo {volume} {90}},\ \bibinfo {pages} {012805} (\bibinfo
  {year} {2014})}\BibitemShut {NoStop}%
\bibitem [{\citenamefont {Karrer}\ \emph {et~al.}(2008)\citenamefont {Karrer},
  \citenamefont {Levina},\ and\ \citenamefont {Newman}}]{Karrer2008}%
  \BibitemOpen
  \bibfield  {author} {\bibinfo {author} {\bibfnamefont {B.}~\bibnamefont
  {Karrer}}, \bibinfo {author} {\bibfnamefont {E.}~\bibnamefont {Levina}}, \
  and\ \bibinfo {author} {\bibfnamefont {M.~E.}\ \bibnamefont {Newman}},\
  }\href@noop {} {\bibfield  {journal} {\bibinfo  {journal} {Physical Review
  E}\ }\textbf {\bibinfo {volume} {77}},\ \bibinfo {pages} {046119} (\bibinfo
  {year} {2008})}\BibitemShut {NoStop}%
\bibitem [{\citenamefont {Lancichinetti}\ \emph {et~al.}(2010)\citenamefont
  {Lancichinetti}, \citenamefont {Radicchi},\ and\ \citenamefont
  {Ramasco}}]{Lancichinetti2010}%
  \BibitemOpen
  \bibfield  {author} {\bibinfo {author} {\bibfnamefont {A.}~\bibnamefont
  {Lancichinetti}}, \bibinfo {author} {\bibfnamefont {F.}~\bibnamefont
  {Radicchi}}, \ and\ \bibinfo {author} {\bibfnamefont {J.~J.}\ \bibnamefont
  {Ramasco}},\ }\href@noop {} {\bibfield  {journal} {\bibinfo  {journal}
  {Physical Review E}\ }\textbf {\bibinfo {volume} {81}},\ \bibinfo {pages}
  {046110} (\bibinfo {year} {2010})}\BibitemShut {NoStop}%
\bibitem [{\citenamefont {Lancichinetti}\ \emph {et~al.}(2011)\citenamefont
  {Lancichinetti}, \citenamefont {Radicchi}, \citenamefont {Ramasco},\ and\
  \citenamefont {Fortunato}}]{Lancichinetti2011b}%
  \BibitemOpen
  \bibfield  {author} {\bibinfo {author} {\bibfnamefont {A.}~\bibnamefont
  {Lancichinetti}}, \bibinfo {author} {\bibfnamefont {F.}~\bibnamefont
  {Radicchi}}, \bibinfo {author} {\bibfnamefont {J.~J.}\ \bibnamefont
  {Ramasco}}, \ and\ \bibinfo {author} {\bibfnamefont {S.}~\bibnamefont
  {Fortunato}},\ }\href@noop {} {\bibfield  {journal} {\bibinfo  {journal}
  {PloS one}\ }\textbf {\bibinfo {volume} {6}},\ \bibinfo {pages} {e18961}
  (\bibinfo {year} {2011})}\BibitemShut {NoStop}%
\bibitem [{\citenamefont {Tumminello}\ \emph {et~al.}(2011)\citenamefont
  {Tumminello}, \citenamefont {Micciche}, \citenamefont {Lillo}, \citenamefont
  {Piilo},\ and\ \citenamefont {Mantegna}}]{Tumminello2011}%
  \BibitemOpen
  \bibfield  {author} {\bibinfo {author} {\bibfnamefont {M.}~\bibnamefont
  {Tumminello}}, \bibinfo {author} {\bibfnamefont {S.}~\bibnamefont
  {Micciche}}, \bibinfo {author} {\bibfnamefont {F.}~\bibnamefont {Lillo}},
  \bibinfo {author} {\bibfnamefont {J.}~\bibnamefont {Piilo}}, \ and\ \bibinfo
  {author} {\bibfnamefont {R.~N.}\ \bibnamefont {Mantegna}},\ }\href@noop {}
  {\bibfield  {journal} {\bibinfo  {journal} {PloS one}\ }\textbf {\bibinfo
  {volume} {6}},\ \bibinfo {pages} {e17994} (\bibinfo {year}
  {2011})}\BibitemShut {NoStop}%
\bibitem [{\citenamefont {Newman}(2004)}]{Newman2004b}%
  \BibitemOpen
  \bibfield  {author} {\bibinfo {author} {\bibfnamefont {M.~E.}\ \bibnamefont
  {Newman}},\ }\href@noop {} {\bibfield  {journal} {\bibinfo  {journal}
  {Physical review E}\ }\textbf {\bibinfo {volume} {70}},\ \bibinfo {pages}
  {056131} (\bibinfo {year} {2004})}\BibitemShut {NoStop}%
\bibitem [{\citenamefont {Serrano}\ \emph {et~al.}(2009)\citenamefont
  {Serrano}, \citenamefont {Bogun{\'a}},\ and\ \citenamefont
  {Vespignani}}]{Serrano2009}%
  \BibitemOpen
  \bibfield  {author} {\bibinfo {author} {\bibfnamefont {M.~{\'A}.}\
  \bibnamefont {Serrano}}, \bibinfo {author} {\bibfnamefont {M.}~\bibnamefont
  {Bogun{\'a}}}, \ and\ \bibinfo {author} {\bibfnamefont {A.}~\bibnamefont
  {Vespignani}},\ }\href@noop {} {\bibfield  {journal} {\bibinfo  {journal}
  {Proceedings of the national academy of sciences}\ }\textbf {\bibinfo
  {volume} {106}},\ \bibinfo {pages} {6483} (\bibinfo {year}
  {2009})}\BibitemShut {NoStop}%
\bibitem [{\citenamefont {Hatzopoulos}\ \emph {et~al.}(2015)\citenamefont
  {Hatzopoulos}, \citenamefont {Iori}, \citenamefont {Mantegna}, \citenamefont
  {Miccich{\`e}},\ and\ \citenamefont {Tumminello}}]{Hatzopoulos2015}%
  \BibitemOpen
  \bibfield  {author} {\bibinfo {author} {\bibfnamefont {V.}~\bibnamefont
  {Hatzopoulos}}, \bibinfo {author} {\bibfnamefont {G.}~\bibnamefont {Iori}},
  \bibinfo {author} {\bibfnamefont {R.~N.}\ \bibnamefont {Mantegna}}, \bibinfo
  {author} {\bibfnamefont {S.}~\bibnamefont {Miccich{\`e}}}, \ and\ \bibinfo
  {author} {\bibfnamefont {M.}~\bibnamefont {Tumminello}},\ }\href@noop {}
  {\bibfield  {journal} {\bibinfo  {journal} {Quantitative Finance}\ }\textbf
  {\bibinfo {volume} {15}},\ \bibinfo {pages} {693} (\bibinfo {year}
  {2015})}\BibitemShut {NoStop}%
\bibitem [{\citenamefont {Saracco}\ \emph {et~al.}(2016)\citenamefont
  {Saracco}, \citenamefont {Di~Clemente}, \citenamefont {Gabrielli},\ and\
  \citenamefont {Squartini}}]{Saracco2016}%
  \BibitemOpen
  \bibfield  {author} {\bibinfo {author} {\bibfnamefont {F.}~\bibnamefont
  {Saracco}}, \bibinfo {author} {\bibfnamefont {R.}~\bibnamefont
  {Di~Clemente}}, \bibinfo {author} {\bibfnamefont {A.}~\bibnamefont
  {Gabrielli}}, \ and\ \bibinfo {author} {\bibfnamefont {T.}~\bibnamefont
  {Squartini}},\ }\href@noop {} {\bibfield  {journal} {\bibinfo  {journal}
  {arXiv preprint arXiv:1607.02481}\ } (\bibinfo {year} {2016})}\BibitemShut
  {NoStop}%
\bibitem [{\citenamefont {Gualdi}\ \emph {et~al.}(2016)\citenamefont {Gualdi},
  \citenamefont {Cimini}, \citenamefont {Primicerio}, \citenamefont
  {Di~Clemente}, \citenamefont {Challet} \emph {et~al.}}]{Gualdi2016}%
  \BibitemOpen
  \bibfield  {author} {\bibinfo {author} {\bibfnamefont {S.}~\bibnamefont
  {Gualdi}}, \bibinfo {author} {\bibfnamefont {G.}~\bibnamefont {Cimini}},
  \bibinfo {author} {\bibfnamefont {K.}~\bibnamefont {Primicerio}}, \bibinfo
  {author} {\bibfnamefont {R.}~\bibnamefont {Di~Clemente}}, \bibinfo {author}
  {\bibfnamefont {D.}~\bibnamefont {Challet}},  \emph {et~al.},\ }\href@noop {}
  {\bibfield  {journal} {\bibinfo  {journal} {arXiv preprint arXiv:1603.05914}\
  } (\bibinfo {year} {2016})}\BibitemShut {NoStop}%
\bibitem [{\citenamefont {Hochberg}\ and\ \citenamefont
  {Tamhane}(2009)}]{Hochberg1987}%
  \BibitemOpen
  \bibfield  {author} {\bibinfo {author} {\bibfnamefont {Y.}~\bibnamefont
  {Hochberg}}\ and\ \bibinfo {author} {\bibfnamefont {A.~C.}\ \bibnamefont
  {Tamhane}},\ }\href@noop {} {\  (\bibinfo {year} {2009})}\BibitemShut
  {NoStop}%
\bibitem [{\citenamefont {Benjamini}\ and\ \citenamefont
  {Hochberg}(1995)}]{Benjamini1995}%
  \BibitemOpen
  \bibfield  {author} {\bibinfo {author} {\bibfnamefont {Y.}~\bibnamefont
  {Benjamini}}\ and\ \bibinfo {author} {\bibfnamefont {Y.}~\bibnamefont
  {Hochberg}},\ }\href@noop {} {\bibfield  {journal} {\bibinfo  {journal}
  {Journal of the royal statistical society. Series B (Methodological)}\ ,\
  \bibinfo {pages} {289}} (\bibinfo {year} {1995})}\BibitemShut {NoStop}%
\bibitem [{\citenamefont {Rand}(1971)}]{Rand1971}%
  \BibitemOpen
  \bibfield  {author} {\bibinfo {author} {\bibfnamefont {W.~M.}\ \bibnamefont
  {Rand}},\ }\href@noop {} {\bibfield  {journal} {\bibinfo  {journal} {Journal
  of the American Statistical association}\ }\textbf {\bibinfo {volume} {66}},\
  \bibinfo {pages} {846} (\bibinfo {year} {1971})}\BibitemShut {NoStop}%
\bibitem [{\citenamefont {Hubert}\ and\ \citenamefont
  {Arabie}(1985)}]{Hubert1985}%
  \BibitemOpen
  \bibfield  {author} {\bibinfo {author} {\bibfnamefont {L.}~\bibnamefont
  {Hubert}}\ and\ \bibinfo {author} {\bibfnamefont {P.}~\bibnamefont
  {Arabie}},\ }\href@noop {} {\bibfield  {journal} {\bibinfo  {journal}
  {Journal of classification}\ }\textbf {\bibinfo {volume} {2}},\ \bibinfo
  {pages} {193} (\bibinfo {year} {1985})}\BibitemShut {NoStop}%
\bibitem [{\citenamefont {Wallace}(1983)}]{Wallace1983}%
  \BibitemOpen
  \bibfield  {author} {\bibinfo {author} {\bibfnamefont {D.~L.}\ \bibnamefont
  {Wallace}},\ }\href@noop {} {\bibfield  {journal} {\bibinfo  {journal}
  {Journal of the American Statistical Association}\ }\textbf {\bibinfo
  {volume} {78}},\ \bibinfo {pages} {569} (\bibinfo {year} {1983})}\BibitemShut
  {NoStop}%
\bibitem [{\citenamefont {Carrico}\ \emph {et~al.}(2006)\citenamefont
  {Carrico}, \citenamefont {Silva-Costa}, \citenamefont {Melo-Cristino},
  \citenamefont {Pinto}, \citenamefont {De~Lencastre}, \citenamefont
  {Almeida},\ and\ \citenamefont {Ramirez}}]{Carrico2006}%
  \BibitemOpen
  \bibfield  {author} {\bibinfo {author} {\bibfnamefont {J.}~\bibnamefont
  {Carrico}}, \bibinfo {author} {\bibfnamefont {C.}~\bibnamefont
  {Silva-Costa}}, \bibinfo {author} {\bibfnamefont {J.}~\bibnamefont
  {Melo-Cristino}}, \bibinfo {author} {\bibfnamefont {F.}~\bibnamefont
  {Pinto}}, \bibinfo {author} {\bibfnamefont {H.}~\bibnamefont {De~Lencastre}},
  \bibinfo {author} {\bibfnamefont {J.}~\bibnamefont {Almeida}}, \ and\
  \bibinfo {author} {\bibfnamefont {M.}~\bibnamefont {Ramirez}},\ }\href@noop
  {} {\bibfield  {journal} {\bibinfo  {journal} {Journal of Clinical
  Microbiology}\ }\textbf {\bibinfo {volume} {44}},\ \bibinfo {pages} {2524}
  (\bibinfo {year} {2006})}\BibitemShut {NoStop}%
\bibitem [{Bon()}]{Bongiorno2017}%
  \BibitemOpen
  \href@noop {} {}\bibinfo {howpublished}
  {\url{https://github.com/cbongiorno/Bipartite-Tools}}\BibitemShut {NoStop}%
\bibitem [{\citenamefont {Newman}(2001)}]{Newman2001}%
  \BibitemOpen
  \bibfield  {author} {\bibinfo {author} {\bibfnamefont {M.~E.}\ \bibnamefont
  {Newman}},\ }\href@noop {} {\bibfield  {journal} {\bibinfo  {journal}
  {Proceedings of the National Academy of Sciences}\ }\textbf {\bibinfo
  {volume} {98}},\ \bibinfo {pages} {404} (\bibinfo {year} {2001})}\BibitemShut
  {NoStop}%
\bibitem [{\citenamefont {Watts}\ and\ \citenamefont
  {Strogatz}(1998)}]{Watts1998}%
  \BibitemOpen
  \bibfield  {author} {\bibinfo {author} {\bibfnamefont {D.~J.}\ \bibnamefont
  {Watts}}\ and\ \bibinfo {author} {\bibfnamefont {S.~H.}\ \bibnamefont
  {Strogatz}},\ }\href@noop {} {\bibfield  {journal} {\bibinfo  {journal}
  {nature}\ }\textbf {\bibinfo {volume} {393}},\ \bibinfo {pages} {440}
  (\bibinfo {year} {1998})}\BibitemShut {NoStop}%
\bibitem [{\citenamefont {Guimer{\`a}}\ \emph {et~al.}(2007)\citenamefont
  {Guimer{\`a}}, \citenamefont {Sales-Pardo},\ and\ \citenamefont
  {Amaral}}]{Guimera2007}%
  \BibitemOpen
  \bibfield  {author} {\bibinfo {author} {\bibfnamefont {R.}~\bibnamefont
  {Guimer{\`a}}}, \bibinfo {author} {\bibfnamefont {M.}~\bibnamefont
  {Sales-Pardo}}, \ and\ \bibinfo {author} {\bibfnamefont {L.~A.~N.}\
  \bibnamefont {Amaral}},\ }\href@noop {} {\bibfield  {journal} {\bibinfo
  {journal} {Physical Review E}\ }\textbf {\bibinfo {volume} {76}},\ \bibinfo
  {pages} {036102} (\bibinfo {year} {2007})}\BibitemShut {NoStop}%
\end{thebibliography}%

\end{document}

\end{document}